\documentclass{article}

\usepackage{set_arxiv}

\usepackage[utf8]{inputenc} 
\usepackage[T1]{fontenc}    
\usepackage{hyperref}       
\usepackage{url}            
\usepackage{caption}
\usepackage{subcaption}
\usepackage{booktabs}       
\usepackage{amsmath}
\usepackage{mathtools}
\usepackage{amssymb}
\usepackage{nicefrac}       
\usepackage{microtype}      
\usepackage{lipsum}		
\usepackage{graphicx}
\usepackage{doi}
\usepackage[numbers]{natbib}
\usepackage{bookmark}

\title{Site-Specific Ground Motion Generative Model for Crustal Earthquakes in Japan Based on Generative Adversarial Networks\thanks{Preprint submitted to arXiv.}}

\author{\href{https://orcid.org/0000-0002-8929-9453}{\includegraphics[scale=0.06]{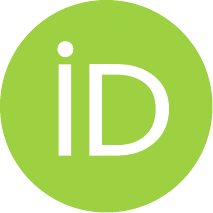}\hspace{1mm}Yuma Matsumoto} \\
    Department of Architecture \\
    Graduate School of Engineering \\
    The University of Tokyo\\
    JSPS Research Fellow \\
    Tokyo, Japan \\
	\And
	\href{https://orcid.org/0000-0001-9589-4630}{\includegraphics[scale=0.06]{orcid.pdf}\hspace{1mm}Taro Yaoyama} \\
    Department of Architecture \\
    Graduate School of Engineering \\
	The University of Tokyo\\
	Tokyo, Japan \\
    \And
    \href{https://orcid.org/0009-0001-3083-9953}{\includegraphics[scale=0.06]{orcid.pdf}\hspace{1mm}Sangwon Lee} \\
    Department of Architecture \\
    Graduate School of Engineering \\
	The University of Tokyo\\
	Tokyo, Japan \\
    \And
	\href{https://orcid.org/0009-0005-4349-7592}{\includegraphics[scale=0.06]{orcid.pdf}\hspace{1mm}Takenori Hida} \\
    Major in Urban and Civil Engineering \\
    Graduate School of Science and Engineering \\
    Ibaraki University \\
	Ibaraki, Japan \\
    \And
    \href{https://orcid.org/0000-0003-3522-7101}{\includegraphics[scale=0.06]{orcid.pdf}\hspace{1mm}Tatsuya Itoi} \\
    Department of Architecture \\
    Graduate School of Engineering \\
	The University of Tokyo\\
	Tokyo, Japan \\
}

\begin{document}
\maketitle

\begin{abstract}
  We develop a site-specific ground-motion model (GMM) for crustal earthquakes in Japan that can directly model the probability distribution of ground motion acceleration time histories based on generative adversarial networks (GANs). The proposed model can generate ground motions conditioned on moment magnitude, rupture distance, and detailed site conditions defined by the average shear-wave velocity in the top 5 m, 10 m, and 20 m ($V_{\mathrm{S}5}$, $V_{\mathrm{S}10}$, $V_{\mathrm{S}20}$) and the depth to shear-wave velocities of 1.0 km/s and 1.4 km/s ($Z_{1.0}$, $Z_{1.4}$). We construct the neural networks based on styleGAN2 and introduce a novel neural network architecture to generate ground motions considering the effect of source, path, and such detailed site conditions. 5\% damped spectral acceleration of ground motions generated by the proposed GMM is consistent with empirical GMMs in terms of magnitude and distance scaling. The proposed GMM can also generate ground motions accounting for the shear-wave velocity profiles of surface soil with different magnitudes and distances, and represent characteristic that are not explained solely by $V_{\mathrm{S}30}$.
\end{abstract}

\section{Introduction}
Ground-motion models (GMMs) are a critical component in probabilistic seismic hazard analysis (PSHA) \cite{Cornel1971}, \cite{Baker2021seismic}, serving as significant inputs for earthquake engineering. The GMM can evaluate the median as well as the variability of ground motions at a specific site, as functions of source characteristics, propagation path effects, and site conditions. In Japan, various empirical GMMs have been constructed for crustal earthquakes, subduction zone earthquakes, or both (e.g., \cite{FukushimaNew1990}, \cite{Molas1995}, \cite{Si1999}, \cite{Nishimura2003}, \cite{Kanno2006}, \cite{Zhao2006Attenuation}, \cite{Morikawa2013}, \cite{Ghofrani2014}, \cite{Zhao2016cr}, \cite{Si2022Development}).
These GMMs were developed using the abundant strong motion data recorded by networks such as K-NET and KiK-net \cite{KNET}.
Some GMMs have been constructed using a global database, while introducing a region-dependent term.
This approach allows for the development of GMMs with the vast amount of data obtained worldwide, while also taking regional characteristics into consideration.
For instance, in the Next Generation Attenuation-West2 (NGA-West2) project \cite{Bozorgnia2014NGA}, a dataset for crustal earthquakes was compiled \cite{Ancheta2014nga}
and GMMs that can be used in Japan were developed \cite{abrahamson2014summary}, \cite{boore2014nga}, \cite{campbell2014nga}, \cite{chiou2014update}.
For subduction zone earthquakes, in NGA-Subduction (NGA-Sub) program \cite{Bozorgnia2022NGA}, GMMs were developed in the same manners \cite{abrahamson2022summary}, \cite{Parker2022NGA}, \cite{Kuehn2023regionalized}.

The aforementioned GMMs were constructed for peak ground acceleration (PGA), peak ground velocity (PGV), and spectral acceleration.
Consequently, hazard curves for these ground motion intensity measures are obtained by PSHA.
However, in recent years, it has become common to conduct non-linear dynamic response analysis using acceleration time histories of ground motion as input \cite{Vamvatsikos2002}
for detailed risk assessment of structures \cite{FEMAp58}.
One approach to synthesizing ground motion time histories using GMMs for dynamic response analysis involves the use of stochastic ground motion models (SGMMs) (e.g., \cite{Kameda1994probabilistic}, \cite{Rezaeian2008}, \cite{Rezaeian2010}).
In the SGMMs, ground motions are described by a non-stationary stochastic model whose parameters are related to the source, path, and site conditions, and
uncertainties in the model parameters are accounted for to represent the variability of the ground motions under the given conditions.
Another approach involves the selection and scaling of ground motions (e.g., \cite{Naeim2004}, \cite{Bommer2004Use}, \cite{Baker2006}).
Ground motions are selected from compiled observed record databases to match the target response spectrum under given conditions.
Although this method is widely used in earthquake engineering, it does not always ensure a sufficient number of records for a specific source, path, and site conditions.

In this study, we utilize a deep generative model to develop a GMM capable of directly modeling the ground motion time history data.
Deep generative models are probabilistic models that employ deep learning techniques.
A key concept of a deep generative model is its ability to capture the inherent distribution of the learned data, and to generate new data that follows this learned distribution.
In other words, instead of simply replicating the learned data, a generative model can generate a set of new data that is statistically similar to the original by capturing its underlying probability distribution \cite{wang2024pivoting}.
It is generally known that deep generative models, which consist of neural networks with many hidden layers, are capable of learning high-dimensional and complex probability distributions \cite{Ruthotto2021}.
By applying a deep generative model for ground motion time history data, it is expected to be possible to construct probabilistic models for such high-dimensional data,
which has been difficult with existing empirical GMMs and SGMMs.
In this study, we refer to such a deep generative model-based GMM as a ground motion generative model (GMGM).
The GMGM could become one option for the application of GMMs in earthquake engineering, such as in dynamic response analysis.

Several studies have examined the application of deep generative models for ground motions.
Esfahani et al. \cite{Esfahani2021Exploring} utilized an autoencoder on the Fourier amplitude spectra (FAS) of ground motions to estimate
the minimum number of predictor variables that is required for a GMM. They also used the trained autoencoder to generate FAS for specific magnitudes and distances.
Among the studies that applied deep generative models for ground motion time histories, generative adversarial networks (GANs) \cite{Goodfellow2014} have been widely adopted.
Wang et al. \cite{Wang2019segam} and Li et al. \cite{li2020seismic2} examined data augmentation by applying GANs to the generation of ground motion time histories, targeting applications in earthquake detection problems.
Similarly, Li et al. \cite{li2020seismic1} and Wang et al. \cite{wang2021seismogen} have conducted studies on data augmentation for ground motions using a technique
called conditional GANs (cGAN) \cite{mirza2014conditional} that can specify the generated data with some condition labels.
Gatti and Clouteau \cite{gatti2020towards} proposed a method for generating ground motions up to high-frequency components by combining physics-based simulation methods and GANs. 
Grijalva et al. \cite{grijalva2021eseismic} applied GANs to the FAS of ground motions obtained in volcanic events.
Matinfar et al. \cite{matinfar2023deep} trained GANs on wavelet-transformed ground motions and developed a method to generate ground motions matching a target response spectrum.
Matsumoto et al. \cite{matsumoto2023fundamental} trained GANs for ground motion time histories, and demonstrated that the trained GANs model
could adequately approximate the distribution of observed record database.
Addressing the characteristics of source, path, and site conditions, similar to GMMs,
Florez et al. \cite{Florez2022} demonstrated the capability of cGAN in generating ground motion time histories conditioned on magnitude, distance, and
the average shear-wave velocity in the top 30 meters ($V_{\mathrm{S}30}$). Following a similar approach, Esfahani et al. \cite{esfahani2023tfcgan} developed a model named TFCGAN,
which learns the time-frequency domain amplitudes of ground motions conditioned on magnitude, distance, and $V_{\mathrm{S}30}$. 
Additionally, they demonstrated how the time-frequency amplitudes produced by the trained cGAN could be used to retrieve the ground motion time histories using a phase retrieval technique.
Shi et al. \cite{shi2024broadband} employed an extension of GANs, known as the generative adversarial neural operator \cite{rahman2022generative},
to construct a model capable of generating ground motion time histories conditioned on magnitude, distance, $V_{\mathrm{S}30}$, and the style of faulting.

Although various approaches were proposed to apply GANs for ground motion data, most studies have typically described propagation path effects using the source distance and site conditions using $V_{\mathrm{S}30}$.
However, such modeling is based on the ergodic assumption \cite{Anderson1999}, which may lead to an overestimation of variability in ground motions.
Recently, studies on non-ergodic GMMs have been actively conducted to eliminate this ergodic assumption \cite{lavrentiadis2023overview}, and in Japan, site-specific GMMs (e.g., \cite{akaba2024}) and non-ergodic GMMs (e.g., \cite{Sung2024}) have also been developed.
It is equally important to address the ergodic assumption in GMGM.
While the fully non-ergodic GMMs take into account all of the source, path, and site effects,
we initially focus on the site effects only.
A GMGM that accounts for detailed site conditions could have a potential to eliminate parts of the ergodic assumption.

In this study, we propose a site-specific GMGM (SS-GMGM) for crustal earthquakes in Japan,
considering detailed site conditions and utilizing GANs on ground motion time history data.
The proposed SS-GMGM specifies site conditions using five condition labels: the average shear-wave velocities in the top 5 m, 10 m, 20 m ($V_{\mathrm{S}5}$, $V_{\mathrm{S}10}$, $V_{\mathrm{S}20}$), and the depth to the layer with shear-wave velocities of 1.0 km/s and 1.4 km/s ($Z_{1.0}$, $Z_{1.4}$).
By combining these five site conditions with the moment magnitude ($M_W$)
and rupture distance ($R_{RUP}$), the SS-GMGM is trained on ground motion time history data conditioned on a seven-dimensional vector: $[M_W, R_{RUP}, V_{\mathrm{S}5}, V_{\mathrm{S}10}, V_{\mathrm{S}20}, Z_{1.0}, Z_{1.4}]$.
We also propose a novel neural network architecture that can generate ground motion time histories with this seven-dimensional vectors.
The quality and distribution of the generated ground motions from the trained SS-GMGM are evaluated, and the performance of the SS-GMGM is demonstrated by comparing it with existing empirical GMMs.
We also discuss how well the specified site conditions correlate with the generated ground motions.

The structure of this paper is as follows.
Section \ref{sec:training} describes the dataset used for training of the SS-GMGM. 
Section \ref{sec:method} outlines the GANs method and the proposed neural network architectures.
The results of the proposed SS-GMGM are shown in section \ref{sec:results},
and section \ref{sec:conc} presents conclusions and future perspectives based on the detailed site conditions-specified training outcomes.
The program code we used for deep learning is available in a GitHub repository,
which can be accessed at \url{https://github.com/Mat-main-00/ss_gmgm}.

\section{Training Datasets}\label{sec:training}%
\subsection{Data selection and correction}
\begin{figure}[t]
  \centering
  \includegraphics[width=0.5\columnwidth]{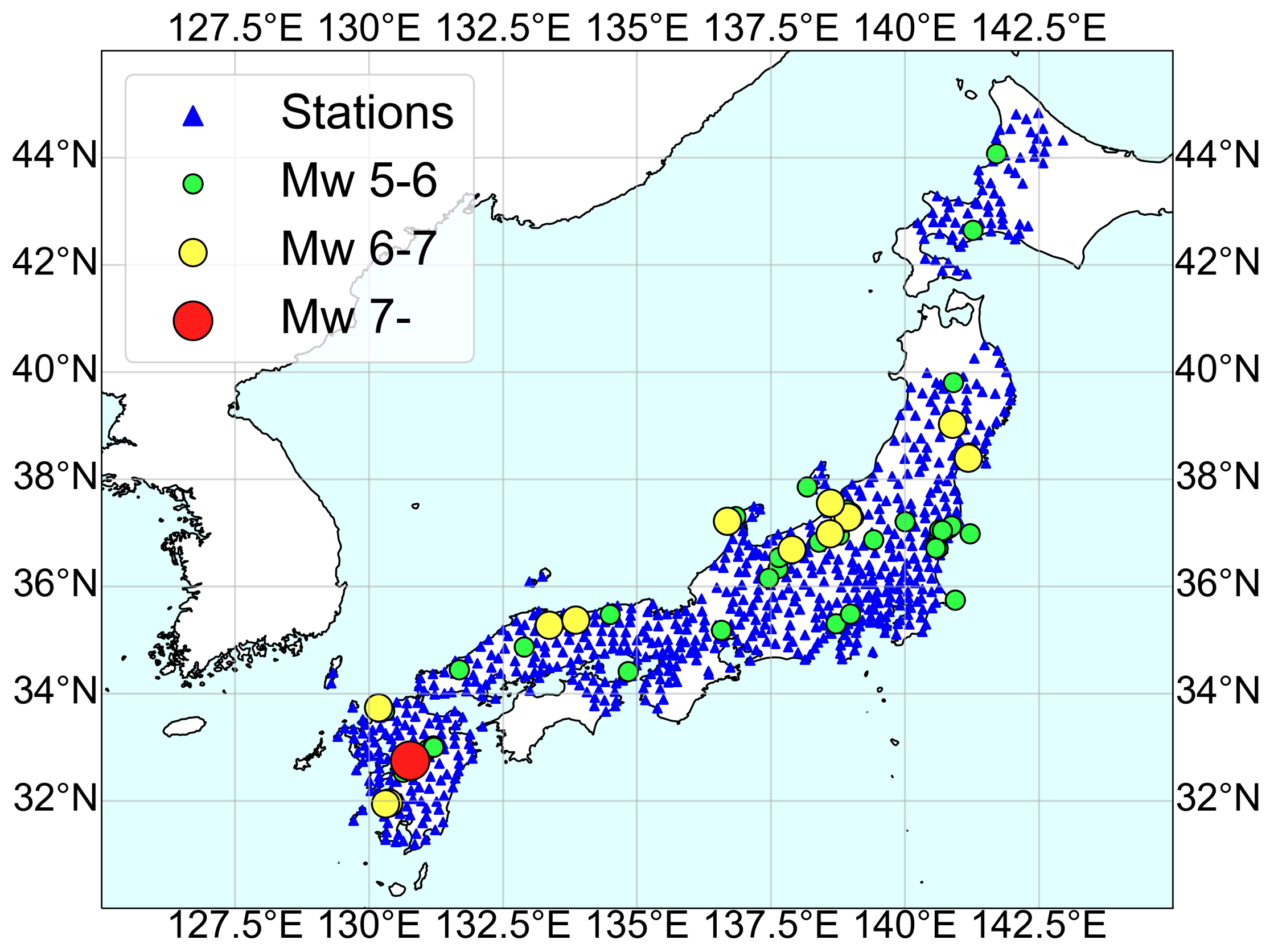}
  \caption{Map of the training dataset showing the locations of the earthquake epicenters (circles) and stations (triangles).}
  \label{fig:knet_source_and_obs_trim}
\end{figure}

\begin{figure}[t]
  \centering
  \includegraphics[width=0.5\columnwidth]{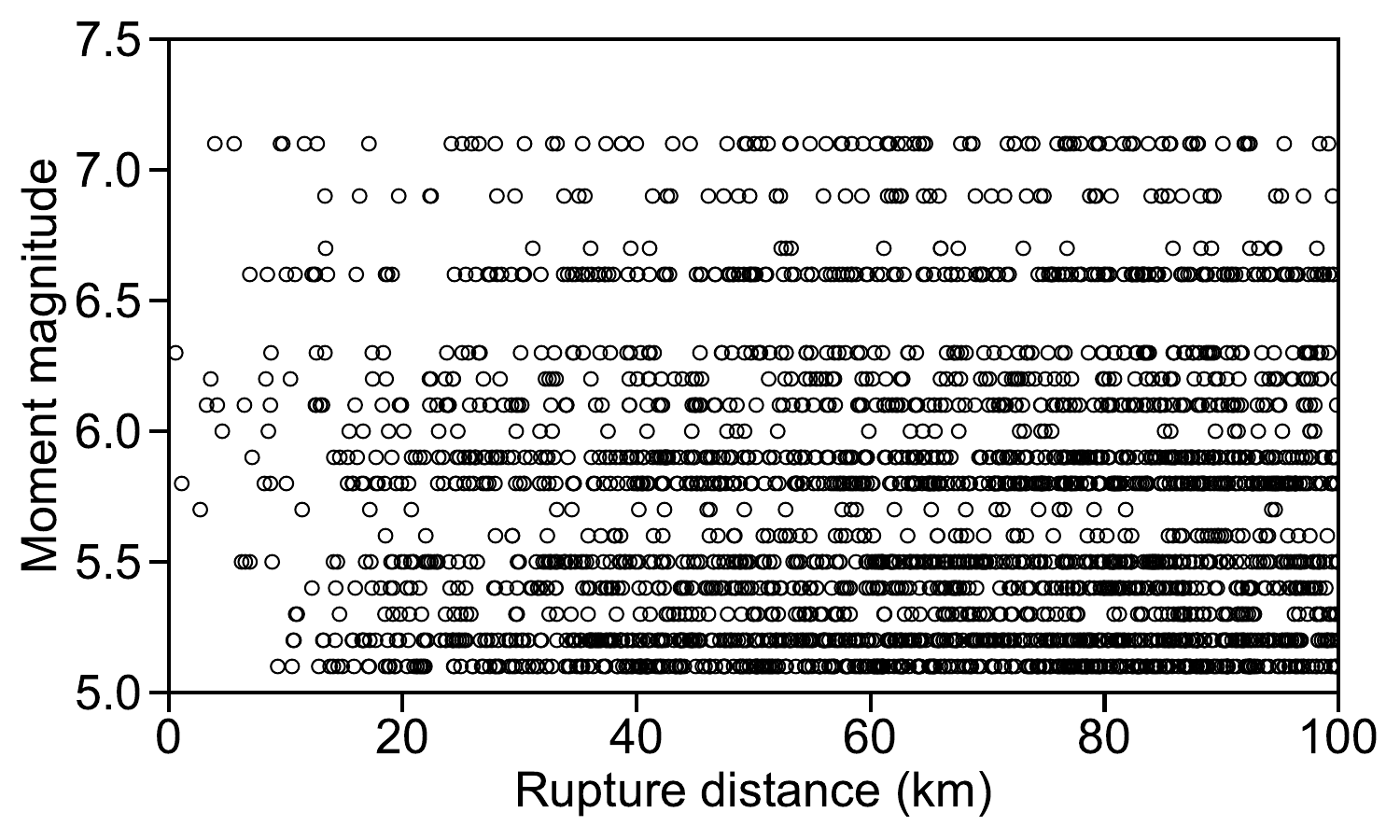}
  \caption{Magnitude-distance distribution of the compiled training dataset.}
  \label{fig:scat_fd_mw}
\end{figure}

To compile the training dataset, we collected observed records from shallow crustal earthquakes in Japan.
The selection criteria for the earthquakes and records are outlined below.
\begin{itemize}
  \item Crustal earthquakes that occurred in the Eurasian plate between 1997 and 2016.
  \item Moment magnitude $M_W > 5$.
  \item Hypocentral depth less than 30 km.
  \item Inclusion of both mainshocks and aftershocks that meet the above criteria.
  \item Observed records at the K-NET stations.
  \item The rupture distance $R_{RUP} \le 100$ km.
  \item Use the two horizontal components of ground motions assuming that they are independent, and ground motions for the rotation angle of 45 degrees are added to augment data.
\end{itemize}
Moment magnitude $M_W$ was determined using the moment tensor solution from the F-net (Full Range Seismograph Network of Japan) database \cite{Fnet},
and the lower threshold of $M_W$ was set to 5.0 referring to the current practices in PSHA in Japan \cite{FUJIWARA2023}.
Rupture distance $R_{RUP}$ was calculated as the shortest distance from the rupture area to the station. When $M_W$ is enough small,
and the earthquake can be considered as a point source,
$R_{RUP}$ was calculated as the hypocenter distance.
The final training dataset consists of 21,696 records from 62 earthquakes.
Each record is a horizontal one-component acceleration time history with 100 Hz sampling.
Details of the selected earthquakes are further described in Matsumoto et al. \cite{matsumoto2023fundamental}.
The locations of the earthquake epicenters and stations are shown in Figure \ref{fig:knet_source_and_obs_trim},
and the magnitude-distance distribution is shown in Figure \ref{fig:scat_fd_mw}.

\subsection{Parameters for site conditions}
The K-NET database provides P-S logging results at one-meter intervals down to a depth of 20 meters from the surface at each station.
Although the SS-GMGM could directly utilize this data as site conditions, we chose to represent surface soil conditions with $V_{\mathrm{S}5}$, $V_{\mathrm{S}10}$, and $V_{\mathrm{S}20}$ considering practical applications.
Therefore, the surface soil is modeled as a three-layer structure in the SS-GMGM.
In many existing GMMs, the surface soil conditions are expressed only by $V_{\mathrm{S}30}$.
For comparative purposes, we also calculated $V_{\mathrm{S}30}$ using the following empirical formula proposed by Kanno et al. \cite{Kanno2006}:
\begin{equation}
  V_{\mathrm{S}30} = 1.13V_{\mathrm{S}20} + 19.5 \label{eq:vs20}
\end{equation}
It is important to note that $V_{\mathrm{S}30}$ is not used as a model parameter in the SS-GMGM but is only used for comparison.
To account for amplification by deep sedimentary layers, Morikawa and Fujiwara \cite{Morikawa2013} used $Z_{1.4}$ in their GMM.
Abrahamson et al. \cite{abrahamson2014summary}, Boore et al. \cite{boore2014nga}, and Chiou and Youngs \cite{chiou2014update} used $Z_{1.0}$ as the parameter to represent that amplification.
Referring to these studies, we use both $Z_{1.0}$ and $Z_{1.4}$ as the site condition parameters of the SS-GMGM.
We obtained the values of $Z_{1.0}$ and $Z_{1.4}$ at the each station included in the complied dataset from the deep subsurface structural model V3.2 provided in Japan Seismic Hazard Information Station (J-SHIS) \cite{jshis}.
It should be noted that the values of $Z_{1.0}$ used in the above three GMMs refer to the depth from the ground surface,
whereas the values in J-SHIS database represent the depth from the engineering bedrock, thus the condition settings are not strictly same.
Furthermore, $Z_{1.0}$ and $Z_{1.4}$ are selected primarily for the purpose of comparison with existing GMMs, and may not necessarily be optimal as the parameters representing the characteristics of ground motions in our dataset.
A future task will be to figure out how to appropriately incorporate the characteristics of deep sedimentary layers into the GMGMs.
Through these procedures, we derived a seven-dimensional vector $[M_W, R_{RUP}, V_{\mathrm{S}5}, V_{\mathrm{S}10}, V_{\mathrm{S}20}, Z_{1.0}, Z_{1.4}]$ as a condition label for each observed record.

\subsection{Data pre-processing}
In the compiled dataset, each ground motion was aligned with respect to the P-wave arrival time,
which wes manually determined by visually inspecting the waveform.
Oscillations caused by different events, such as aftershocks occurring shortly after the mainshock, were identified and removed based on visual inspection of the waveform.
The data length was then set to 7,992 (79.92 seconds) by truncating the end of each record to ensure that regions of large amplitude were retained for almost all observed records.
When the length of the original record was less than 7,992, zeros were appended to the end to equalize the data length across the dataset.
A cosine taper was applied to the final 100 steps, and 100 zeros were appended to the start and finish of the records in order to lessen edge effects and stabilize the learning process.
The sampling frequency was not changed from 100 Hz, and no band-pass filter was applied.
Consequently, ground motion acceleration time histories with a duration of 81.92 seconds (8,192 samples), starting 1.0 second before the onset of the P-wave,
were obtained.

Training of GANs models can be unstable \cite{Arjovsky2017}.
To improve the stability of the learning process, normalization techniques are commonly employed within the neural network architecture \cite{radford2016unsupervised}. 
Therefore, in this study, the waveforms and their amplitudes are learned separately to improve model performance.
Each observed record amplitude was normalized by its PGA, and the PGA value was appended to the corresponding condition label to form an eight-dimensional vector.
Then, normalization was also performed on each element of the condition label vectors, ensuring a mean of zero and a standard deviation of 0.1.

\section{Methods}\label{sec:method}
\subsection{GANs}
GANs consist of two deep neural networks: a generator $G$ and a discriminator $D$.
The generator $G$ receives a noise vector $\boldsymbol{\mathrm{z}}\sim \mathcal{N}(\boldsymbol{0}, \boldsymbol{\mathrm{I}})$ as input to generate new data $G(\boldsymbol{\mathrm{z}})$ (referred to as generated data).
The discriminator $D$ takes both observed data $\boldsymbol{\mathrm{x}}$ and generated data $G(\boldsymbol{\mathrm{z}})$ as input and estimate the probability that the input is observed data.
GANs are trained through an iterative process where the generator and the discriminator are alternately updated. 
The discriminator is trained to correctly distinguish between the observed and generated data, whereas the generator attempts to produce data that the discriminator will mistakenly identify as observed data.
When appropriately trained through such procedures, the discriminator is known to accurately capture the distribution of the learned dataset, enabling the generator to generate realistic new data that follows this learned distribution \cite{Ruthotto2021}.

\begin{figure}[t]
  \centering
  \includegraphics[width=0.4\columnwidth]{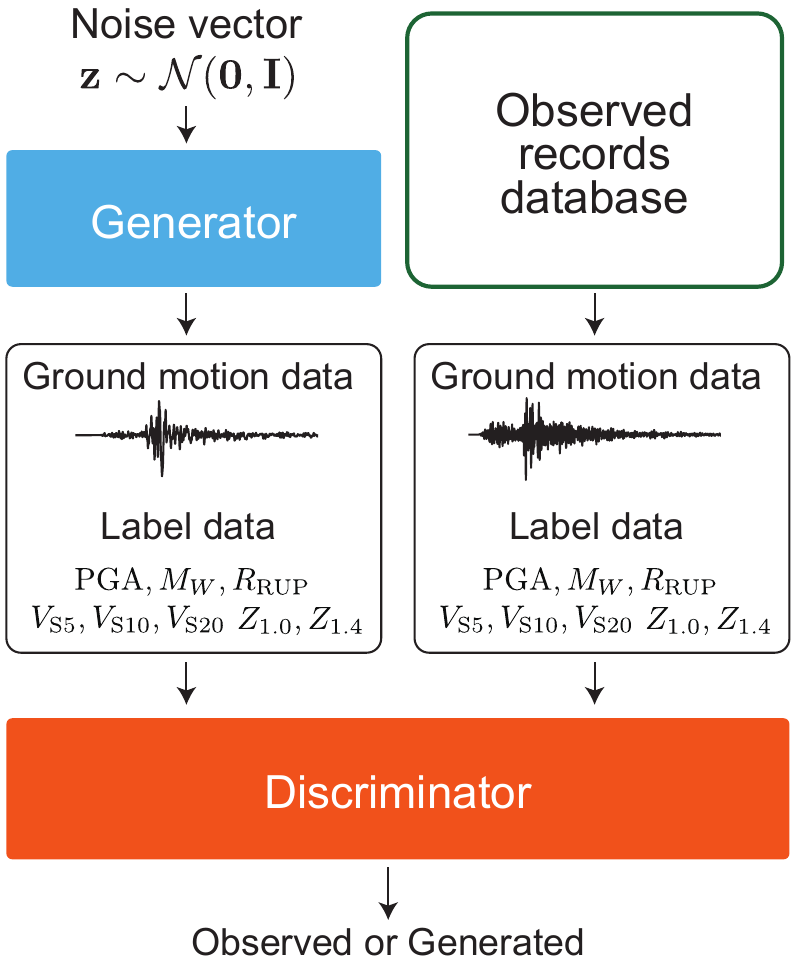}
  \caption{Diagram of overall architecture of the SS-GMGM.}
  \label{fig:overview_gans}
\end{figure}%

In our previous study \cite{matsumoto2023fundamental}, we used a GANs model known as Wasserstein GAN with gradient penalty (WGAN-GP) \cite{Gulrajani2017}. However, in this study,
we propose a model based on styleGAN2 \cite{karras2020analyzing}, which has achieved higher quality data generation, to construct the SS-GMGM.
A key feature of styleGAN2 is the architecture of its generator.
The generator $G$ of styleGAN2 is composed of two neural networks: the mapping network $f$ and the synthesis network $g$.
The mapping network takes the noise vector $\boldsymbol{\mathrm{z}}$ as input and output a vector $\boldsymbol{\mathrm{w}}$, referred to as the intermediate latent variable.
The synthesis network then takes this intermediate latent variable as input and output generated data $g(\boldsymbol{\mathrm{w}})$ (equivalent to $G(\boldsymbol{\mathrm{z}})$).
In conventional GANs, the generator learns the correspondence between the noise vector $\boldsymbol{\mathrm{z}}$ that follows a normal distribution and the observed data $\boldsymbol{\mathrm{x}}$.
However, observed data $\boldsymbol{\mathrm{x}}$ are typically not distributed according to a normal distribution.
By transforming the noise vector $\boldsymbol{\mathrm{z}}$ into an intermediate latent variable $\boldsymbol{\mathrm{w}}$, the input of the synthesis network  is not sampled according to any fixed distribution, but its sampling density is induced by a learned mapping $ f(\boldsymbol{\mathrm{z}})$. This would make it easier for the synthesis network to learn the relationship between the intermediate latent variables $\boldsymbol{\mathrm{w}}$ and observed data $\boldsymbol{\mathrm{x}}$ \cite{Karras2021}.

Following the method of Karras et al. \cite{karras2020analyzing}, we set the objective function for the generator training as follows:
\begin{equation}
  \min_{\boldsymbol{\theta}} \mathbb{E}_{\boldsymbol{\mathrm{z}}}\left[\log(1 - D(G(\boldsymbol{\mathrm{z}}; \boldsymbol{\theta})))\right]
  + \mathbb{E}_{\boldsymbol{\mathrm{w}}, \boldsymbol{\mathrm{u}}\sim\mathcal{N}(0, \boldsymbol{\mathrm{I}})}\left(\left\|
    \boldsymbol{\mathrm{J}}^\mathrm{T}_{\boldsymbol{\mathrm{w}}}\boldsymbol{\mathrm{u}}
  \right\|_2 - a\right)^2 \label{eq:gen}
\end{equation}
where $\boldsymbol{\theta}$ is the parameters of the generator, $\boldsymbol{\mathrm{J}}_{\boldsymbol{\mathrm{w}}}$
is the Jacobian matrix $\boldsymbol{\mathrm{J}}_{\boldsymbol{\mathrm{w}}} = \partial g(\boldsymbol{\mathrm{w}})/\partial \boldsymbol{\mathrm{w}}$, and $a$ is the constant.
The first term is the logistic loss of the normal GANs \cite{Goodfellow2014}, and the second term is the regularization term.
Similarly, we set the objective function for the discriminator training as follows:
\begin{equation}
  \min_{\boldsymbol{\psi}} -\mathbb{E}_{\boldsymbol{\mathrm{x}}}\left[\log D(\boldsymbol{\mathrm{x}}; \boldsymbol{\psi})\right]
  - \mathbb{E}_{\boldsymbol{\mathrm{z}}}\left[\log(1 - D(G(\boldsymbol{\mathrm{z}}); \boldsymbol{\psi}))\right] + \frac{\gamma}{2}\mathbb{E}_{\boldsymbol{\mathrm{x}}}\left[\left\|
    \nabla D(\boldsymbol{\mathrm{x}};\boldsymbol{\psi})
  \right\|^2\right]
\end{equation}
where $\boldsymbol{\psi}$ is the parameters of the discriminator, and $\gamma$ is the constant.
The first and second terms are the logistic losses of the normal GANs \cite{Goodfellow2014}, and the third term is the regularization term called $R_1$ regularization \cite{Mescheder2018}.

\begin{figure}
  \centering
  \includegraphics[width=\columnwidth]{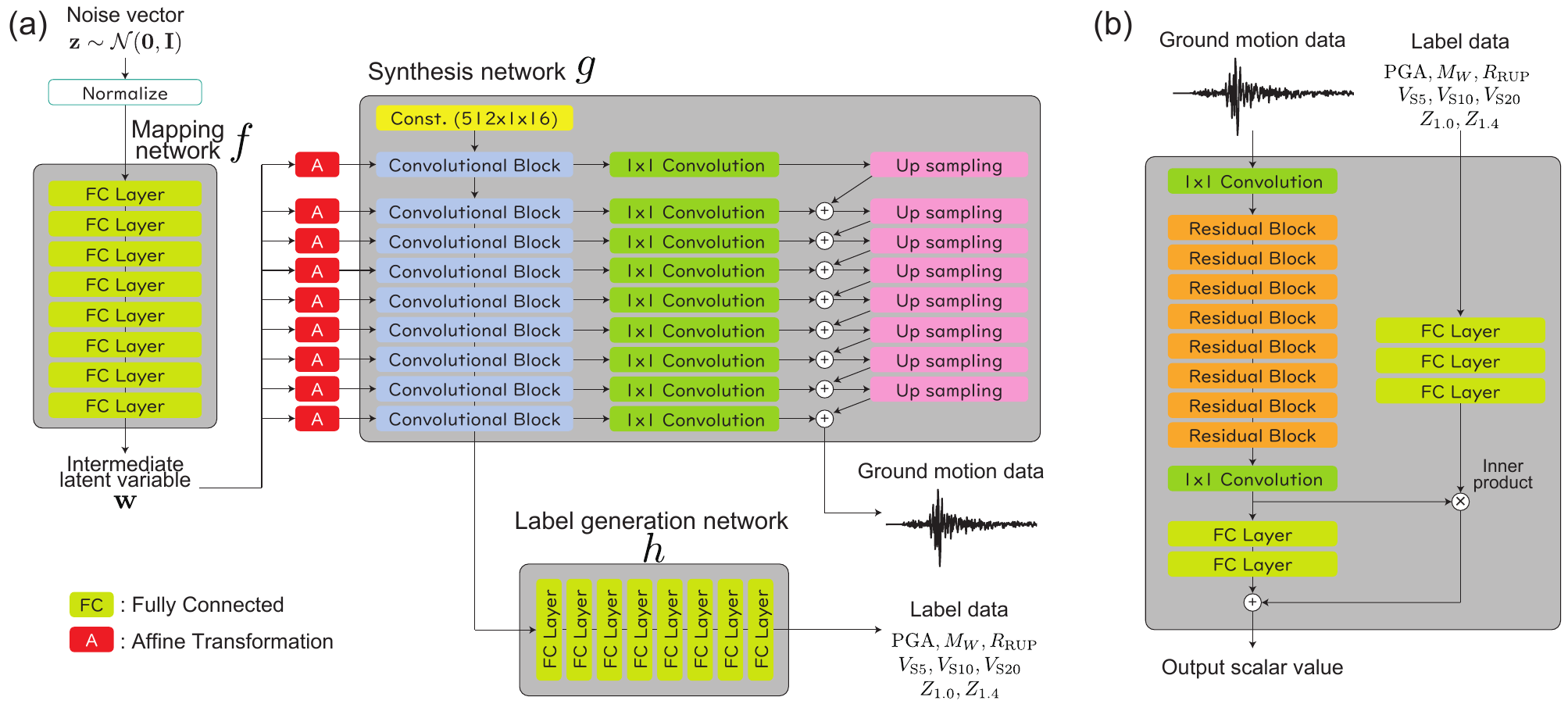}
  \caption{Diagram of the neural network architecture of the proposed SS-GMGM. (a) is the generator, and (b) is the discriminator.}
  \label{fig:generator}
\end{figure}%

\subsection{Proposed neural network architecture of the SS-GMGM}
In this section, we introduce the overview of our newly proposed neural network architecture of the SS-GMGM.
For more detailed information on the neural networks architectures and their parameter settings, please refer to our GitHub repository \url{https://github.com/Mat-main-00/ss_gmgm}.

The overall architecture of the SS-GMGM is shown in Figure \ref{fig:overview_gans}.
The generator takes the noise vector as input and generates a ground motion along with a corresponding condition label.
The discriminator, receiving pairs of ground motion and condition label as input, outputs the probability that the inputs are observed records.
Since the conventional styleGAN2 model cannot handle condition labels,
we made several modifications to the network configurations of Karras et al. \cite{karras2020analyzing}.
An overview of the proposed neural network architecture of the generator is shown in Figure \ref{fig:generator} (a).
A new label generation network $h$ was added to generate the condition labels using the feature maps in the final layer of the synthesis network as input.
The architectures of the mapping network and synthesis network are almost same as the model configuration by Karras et al. \cite{karras2020analyzing},
however, the parameters of the neural networks were modified to fit the data shape of the ground motions.
The mapping network consists of eight fully connected layers, using leaky ReLU \cite{maas2013rectifier} as the activation function.
Skip connection \cite{karras2018progressive} is used to construct the synthesis network, and both ELU \cite{clevert2016fast} and leaky ReLU are used as the activation function.
The label generation network consists of eight fully connected layers, with leaky ReLU employed as the activation function.

An overview of neural network architecture of the discriminator is shown in Figure \ref{fig:generator} (b).
The discriminator receives the ground motion data and label data in different neural networks.
The neural network for the ground motion data was constructed in the same way as the generator, modifying the neural network configurations and parameters of Karras et al. \cite{karras2020analyzing}. The residual network \cite{He2016CVPR} is utilized, and leaky ReLU is applied as the activation function.
We introduced a three fully connected layers for label data referring to a configuration called projection discriminator \cite{miyato2018cgans}.
The outputs of this introduced network are combined with the feature maps of the network for ground motion data by taking inner product.
Through this process, the discriminator is able to comprehend the information from the condition labels in an appropriate manner \cite{miyato2018cgans}.

\begin{figure}
  \centering
  \includegraphics[width=\columnwidth]{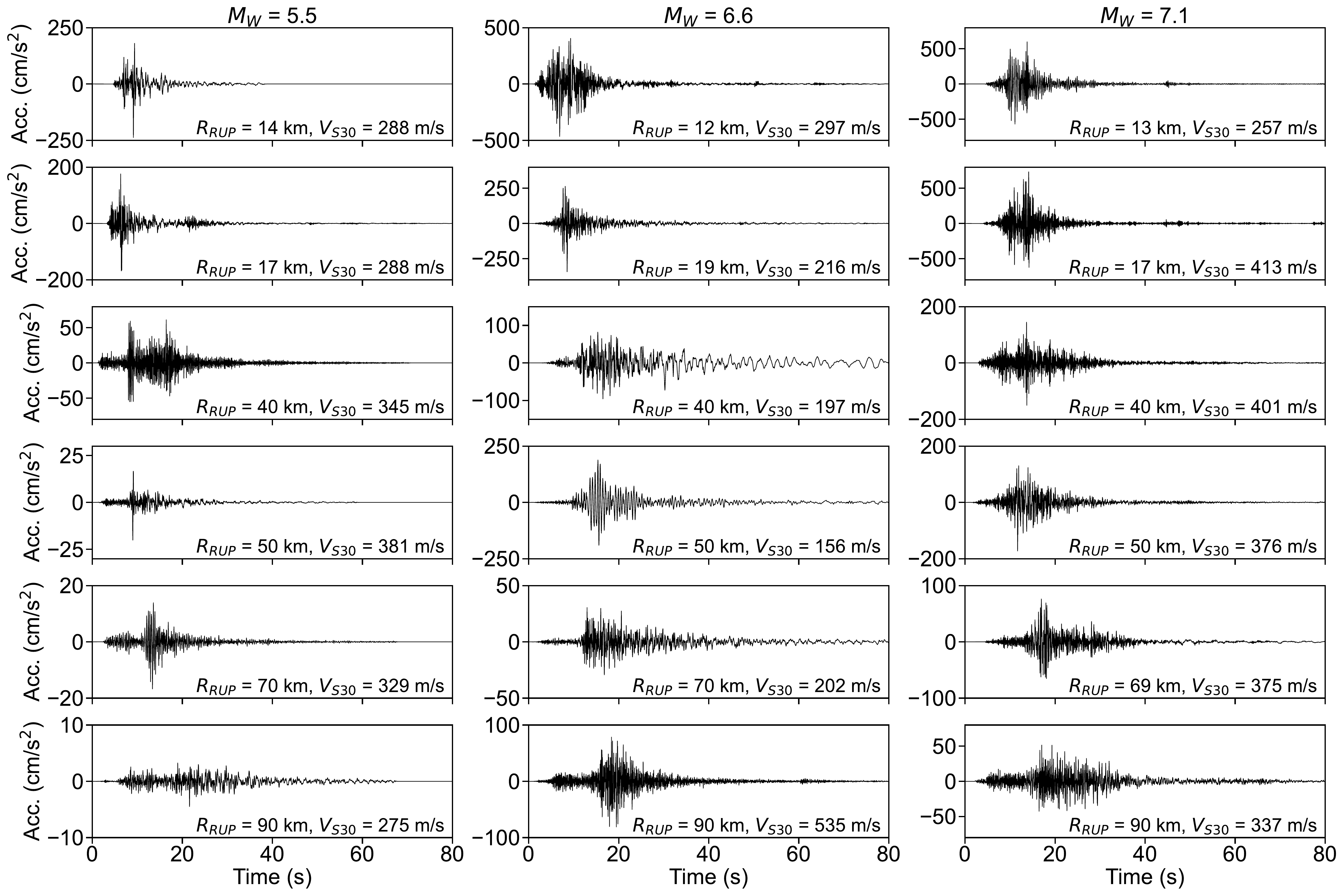}
  \caption{Examples of the ground motion waveforms of observed records.
    The data in each column correspond to the value of $M_W$ shown at the top. Each panel shows the associated $R_{RUP}$ and $V_{\mathrm{S}30}$ values.}
  \label{fig:obs_wave_example}
\end{figure}%

The hyperparameters were determined according to Karras et al. \cite{karras2020analyzing}.
The learning rate was set to 0.002, the batch size was 64, and the dimensions of $\boldsymbol{\mathrm{z}}$ and $\boldsymbol{\mathrm{w}}$ were both set to 512.
Adam \cite{kingma2017adam} was used as the optimization method.
The deep learning and the construction of neural networks were carried out using the Python library PyTorch \cite{Paszke2019}.
For other hyperparameter settings, please refer to our GitHub repository.

\begin{figure}
  \centering
  \includegraphics[width=\columnwidth]{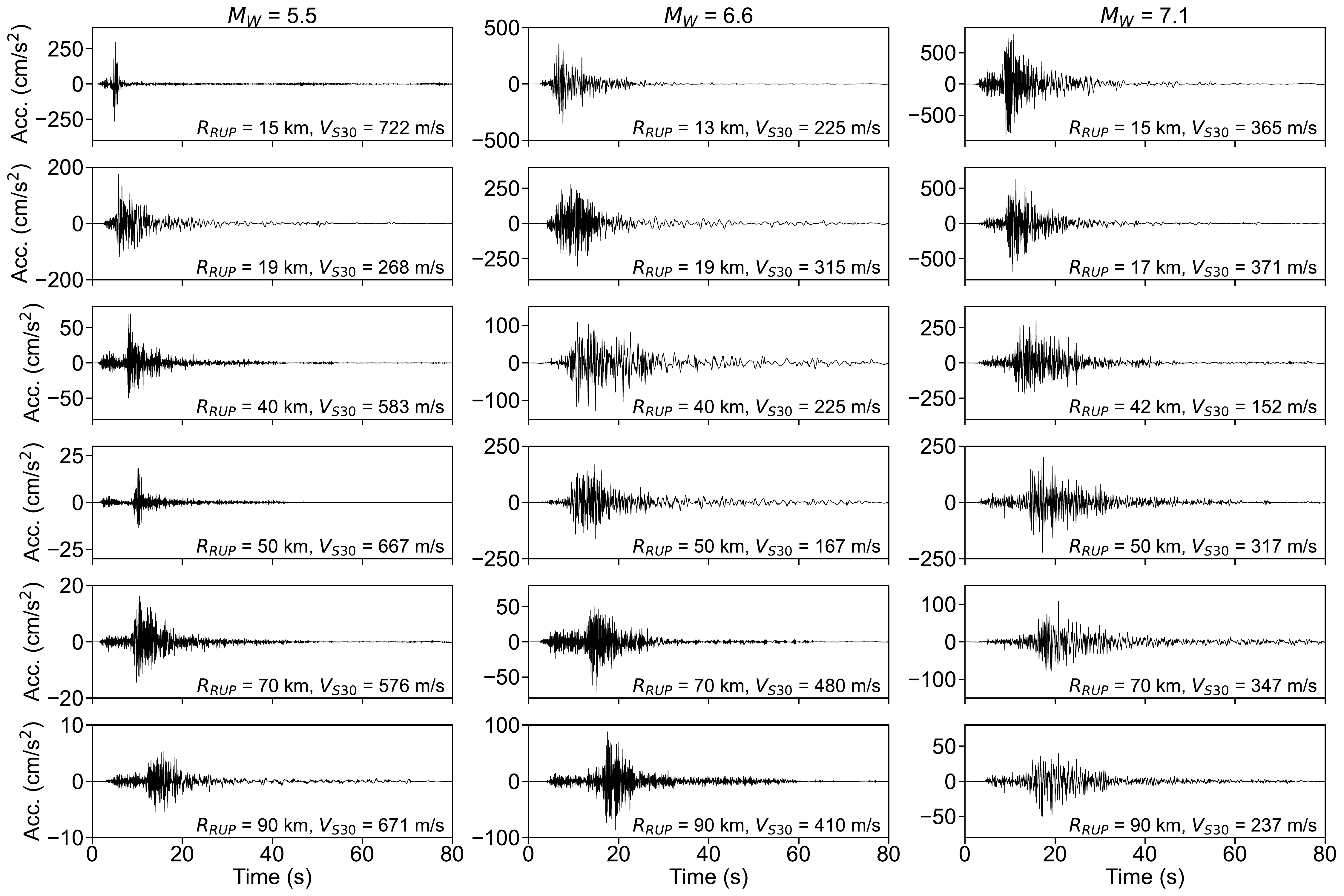}
  \caption{Examples of the ground motion waveforms generated by the SS-GMGM.
    The data in each column correspond to the value of $M_W$ shown at the top. Each panel shows the associated $R_{RUP}$ and $V_{\mathrm{S}30}$ values.}
  \label{fig:gen_wave_example}
\end{figure}%

\section{Analysis Results and Model Evaluation}\label{sec:results}
The SS-GMGM was trained using the compiled dataset, and the trained model generated 100,000 different ground motions and corresponding condition labels.
Several post-processing steps were performed on the each generated ground motion.
Initially, the offset was removed so that mean acceleration becomes zero. Then a fourth-order Butterworth filter was applied
with 0.1 Hz low-frequency cutoff and 20 Hz high-frequency cutoff.
The accelerations in the first and last two seconds were set to zero, similar to the conditions of the training data.
Finally, the amplitudes of each generated ground motion were recovered by multiplying the values of PGA in the corresponding generated condition labels.
In this section, we evaluate the performance of the SS-GMGM using these generated ground motions and condition labels.

\subsection{Generated ground motion waveforms}
We first evaluate the performance of the SS-GMGM by visually checking the generated data.
Figures \ref{fig:obs_wave_example} and \ref{fig:gen_wave_example} show the ground motion waveforms of observed records and generated ground motions for different 
magnitude, distance, and $V_{\mathrm{S}30}$ scenarios, respectively.
$V_{\mathrm{S}30}$ for each generated ground motion was calculated from the generated $V_{\mathrm{S}20}$ using equation \ref{eq:vs20}.
The SS-GMGM appropriately captures waveform characteristics such as the onset of P-waves and S-waves, as well as the envelope shapes.
The amplitude scaling with magnitude and distance is generally well represented, and the relationships between distance and duration, as well as between distance and P-S time, are also appropriate.
Moreover, focusing on the values of $V_{\mathrm{S}30}$, lower frequency ground motions are generated in conditions of soft soil where $V_{\mathrm{S}30} < 300$ m/s, compared to those of stiff soil where $V_{\mathrm{S}30} > 300$ m/s.
In the data of row 3 and column 2 in Figures \ref{fig:obs_wave_example} and \ref{fig:gen_wave_example},
a tendency for subsequent vibrations to have lower frequency components in soft soil condition is also captured.
The more detailed examinations of the correspondence between ground motions and site conditions are conducted in subsection \ref{subsec:ss}.

\subsection{Overall characteristics of generated ground motions}

A key characteristic of the SS-GMGM is its ability to directly learn the probability distribution of ground motion time histories.
Therefore, we examine the distribution of the generated ground motions by comparing it with that of training dataset.
The evaluation is conducted by examining the following five
indices used in Rezaeian and Der Kiureghian \cite{Rezaeian2008}, \cite{Rezaeian2010} to represent the characteristic of ground motions:
\begin{enumerate}
  \item Arias intensity, $I_A$.
  \item Significant/Relative duration of ground motion, $D_{5-95}$.
  \item Significant/Relative duration of ground motion, $D_{5-45}$.
  \item Zero-level crossing rate, $\nu$.
  \item Mean of the rates of negative maxima and positive minima, $\eta$.
\end{enumerate}
Arias intensity \cite{arias1970measure} is a measure of total energy contained in the ground motion and is defined as:
\begin{equation}
  I_A = \frac{\pi}{2g}\int_0^{t_d}a^2(t)\mathrm{d}t
\end{equation}
where $g$ is the gravitational acceleration, $t_d$ is the total duration of ground motion, and $a(t)$ is the acceleration at time $t$.
We set $g = 980.665$ cm/s$^{\text{2}}$.
Significant duration \cite{Bommer1999effective} is widely used as an index for assessing the ground motion duration \cite{Hancock2006}.
The value of $D_{5-95}$ is defined as the time interval required for the cumulative power of the ground motion to reach from 5\% to 95\% of the total cumulative power,
and generally corresponds to the duration of strong motion \cite{Trifunac1975}.
$D_{5-45}$ is also defined as the time interval from 5\% to 45\%, and corresponds to the time at the middle of the strong-shaking phase \cite{Rezaeian2010}.
The Zero-level crossing rate $\nu$ is adopted to characterize the dominant frequency in ground motion \cite{Rezaeian2008},
and we calculated the value of $\nu$ as the mean of zero-level up-crossing rate and zero-level down-crossing rate. 
Negative maxima and positive minima are used to characterize the bandwidth of ground motions \cite{Rezaeian2010}.
It is known that a ground motion with wider bandwidth tends to have larger $\eta$. We defined $\eta$ as the mean of the rates of negative maxima and positive minima.

\begin{figure}
  \centering
  \includegraphics[width=\columnwidth]{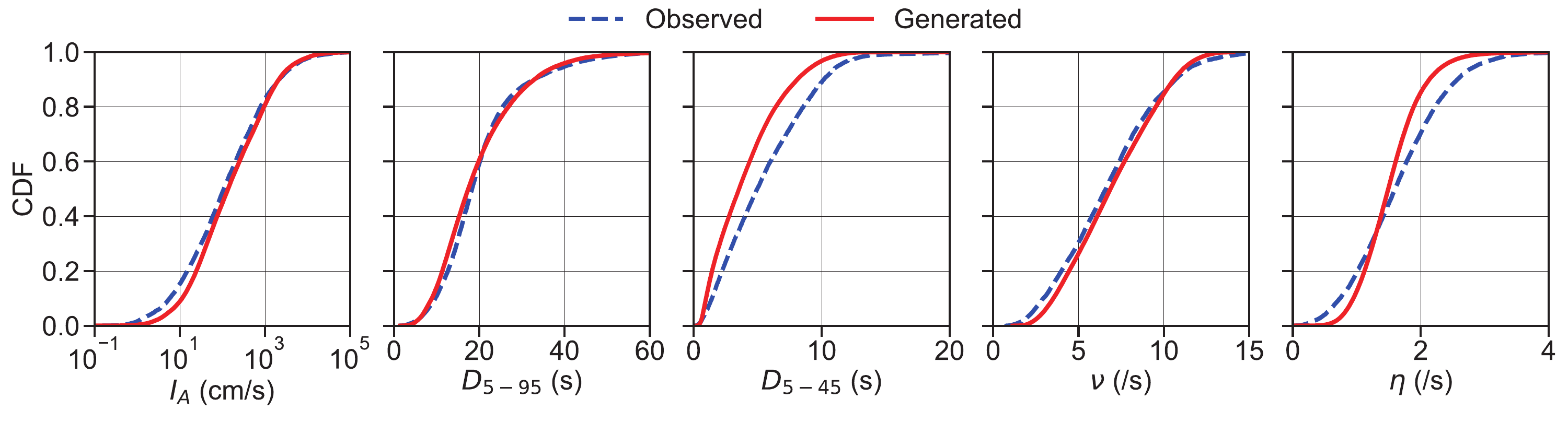}
  \caption{
    Comparison of the cumulative distribution functions (CDF) for five indices of ground motions. The blue dashed line is the CDF of observed records, 
    and red solid line is the CDF of generated ground motions.}
  \label{fig:feature_cdf}
\end{figure}%

\begin{figure}[t]
  \centering
  \includegraphics[width=0.5\columnwidth]{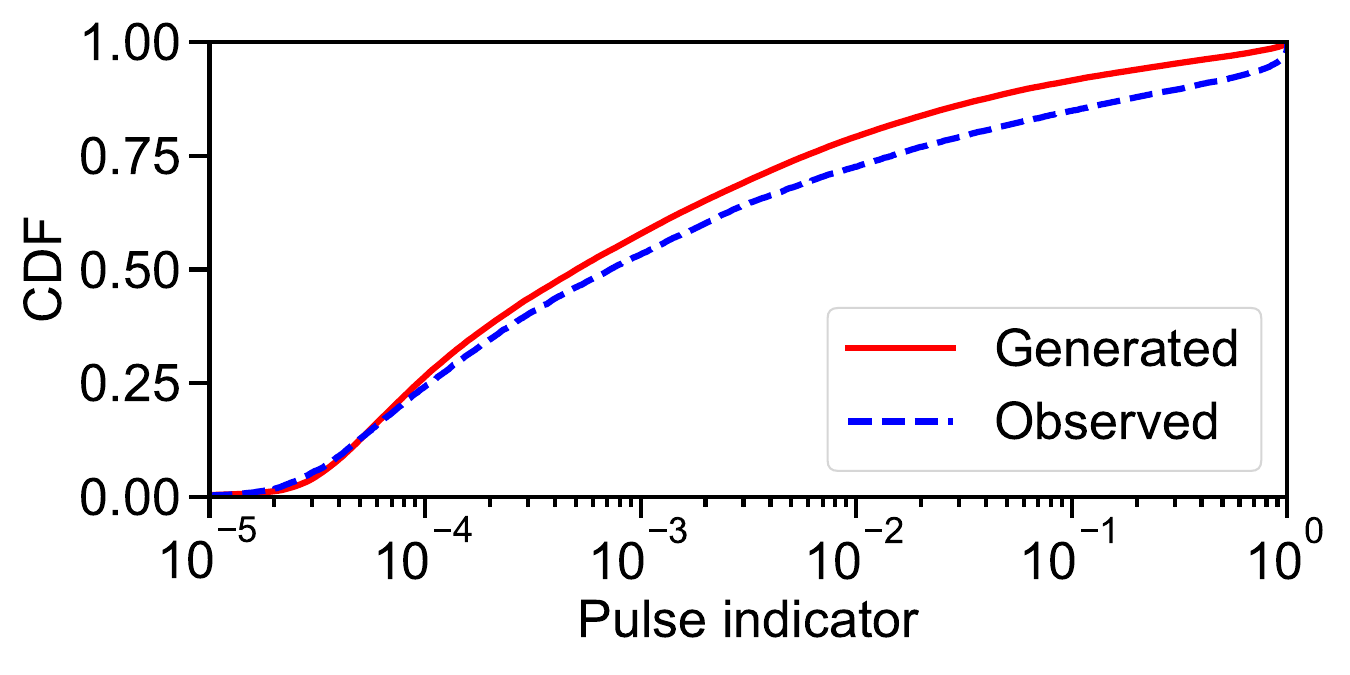}
  \caption{Comparison of the CDF of the pulse indicator values. The blue dashed line represents the observed records, and red solid line represents the generated ground motions.}
  \label{fig:pi_hist}
\end{figure}%

We calculated the values of these five indices for both observed records and generated ground motions.
Note that the values of $\nu$ and $\eta$ were calculated for the vibrations in the time interval corresponding to $D_{5-95}$,
and observed records are filtered with the same Butterworth filter used for post-processing of generated ground motions.
Figure \ref{fig:feature_cdf} shows the comparison results of cumulative distribution function (CDF) of each index.
Although the generated ground motions tend to have a little smaller $D_{5-45}$ values,
the distributions of $I_A$ and $D_{5-95}$ are consistent with that of observed records.
This means that the SS-GMGM captures the temporal characteristic of observed records with reasonable accuracy.
The distribution of $\nu$ also shows a good agreement.
The distribution of $\eta$ indicate that ground motions in the tails of the distribution of observed records are not generated extensively,
however, the overall distribution of generated ground motions is generally matched with that of observed records.
The SS-GMGM captures the overall characteristics of the observed records in terms of both temporal and frequency characteristics,
and is an appropriate probabilistic model of ground motion time histories that
approximates the distribution of learned dataset. 

\begin{figure}
  \centering
  \includegraphics[width=\columnwidth]{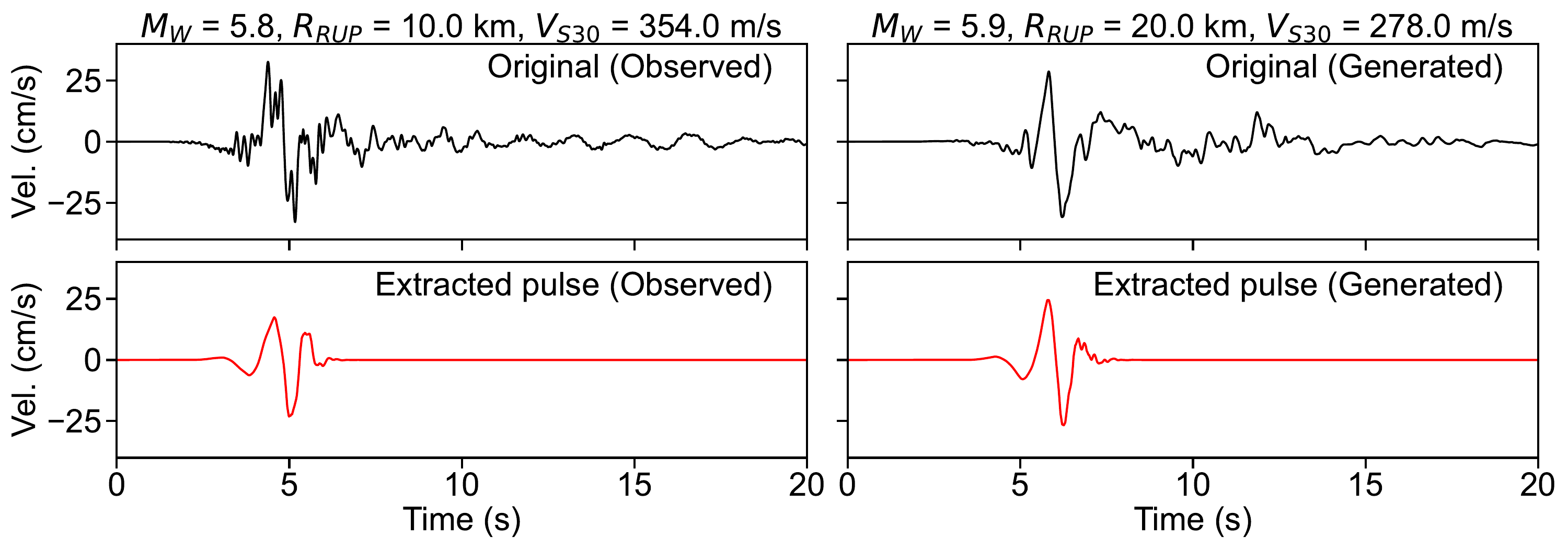}
  \caption{
    Examples of the velocity waveforms and their corresponding extracted pulses. The left column is the observed record and right column is the generated ground motion.
    The scenario $M_W$, $R_{RUP}$, and $V_{\mathrm{S}30}$ of each data is shown at the top of each column.}
  \label{fig:pulse_wave}
\end{figure}%

\subsection{Characteristics of near-field ground motions}
Near-field ground motions containing strong velocity pulse have caused destructive damage to structures (e.g., \cite{Pitarka1998}, \cite{Lin2018}),
and are one of the critical ground motions that should be considered in earthquake engineering.
Following the numerical integration of the generated ground motions, the method of Baker \cite{Baker2007} was applied to classify and extract the velocity pulses.
In the numerical integration of the generated data, a fourth-order Butterworth filter (cutoff frequency 0.2Hz) was used to remove the low frequency components to prevent the noise from them.
It should be noted that while the method of Baker \cite{Baker2007} was designed to be applied to the fault normal components of ground motions,
the generated ground motions by the SS-GMGM do not contain information on the positional relationship with the fault plane. Therefore, it was applied directly to the generated ground motions.

First, following the evaluation in our previous study \cite{matsumoto2023fundamental},
we compare the distribution of following pulse indicator values proposed by Baker \cite{Baker2007}:
\begin{equation}
  \text{Pulse indicator} = \frac{1}{1 + \exp(-23.3 + 14.6v_r + 20.5E_r)}
\end{equation}
where $v_r$ is the PGV of the residual data, which is obtained by subtracting the extracted pulse from original velocity data,
and $E_r$ is the total accumulate power of residual data divided by that of original data.
Pulse-like ground motion tends to have high pulse indicator value.
Figure \ref{fig:pi_hist} shows the CDF of the pulse indicator of observed records and generated ground motions.
Although the proportion of pulse-like ground motion in the generated data is slightly smaller than in the observed records dataset,
the overall trend of the distribution is generally consistent.
Figure \ref{fig:pulse_wave} shows the examples of the original velocity waveforms which were classified as pulse-like ground motion and corresponding extracted pulse waveforms.
A clear velocity pulse is found in generated ground motion with the near-filed setting.
The proposed SS-GMGM is capable of generating ground motions with engineering-significant characteristic.

\begin{figure}
  \centering
  \includegraphics[width=\columnwidth]{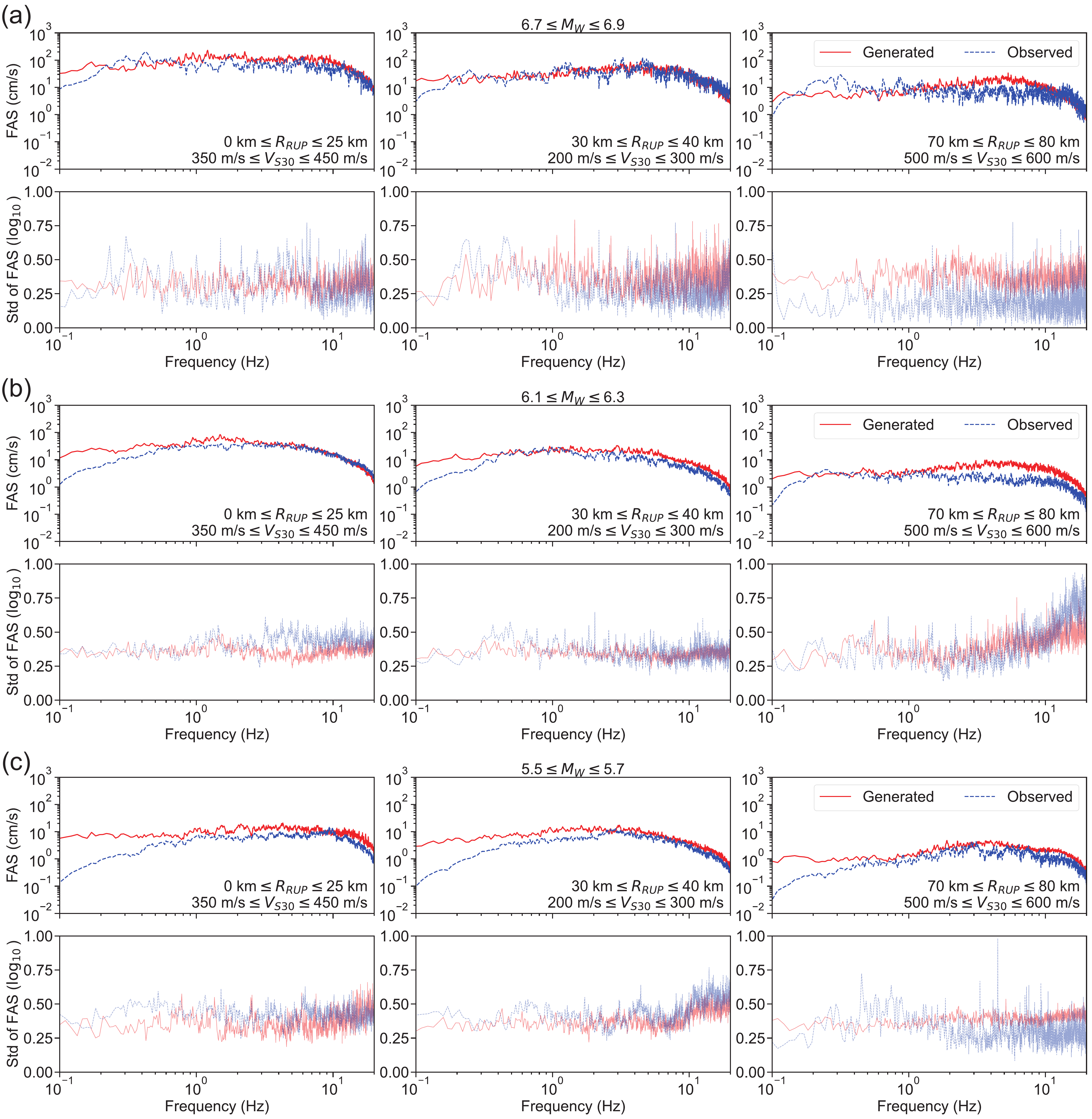}
  \caption{
    Comparison of the logarithmic means and logarithmic standard deviations of the FAS across three different $M_W$ ranges (a-c).
    Panels (a), (b), and (c) have three different $R_{RUP}$ and $V_{\mathrm{S}30}$ scenarios, respectively.
    For each panel, the blue dashed line corresponds to observed records,
    and the red solid line corresponds to generated ground motions.
  }
  \label{fig:fig_famp}
\end{figure}%

\subsection{Statistical evaluation of the FAS}
In this subsection, we conduct statistical performance evaluation of the SS-GMGM.
Following the evaluation method of Florez et al. \cite{Florez2022} and Esfahani et al. \cite{esfahani2023tfcgan},
we compare the distributions of observed and generated FAS with different $M_W$, $R_{RUP}$, and $V_{\mathrm{S}30}$ scenarios.
Figure \ref{fig:fig_famp} compares the logarithmic means and logarithmic standard deviations (std) of the FAS of both observed records and generated ground motions.
The bins of the condition labels are determined so that enough number of observed records are contained.
The generated FAS are generally consistent with observed FAS in frequency range [1, 20] Hz.
In comparisons under relatively small $M_W$ conditions, the generated ground motions contain many low frequency components regardless of $R_{RUP}$ or $V_{\mathrm{S}30}$,
resulting in an overestimation particularly in the region below 0.5 Hz.
On the other hand, in the region with larger magnitudes range $6.7 \le M_W \le 6.9$ (Figure \ref{fig:fig_famp} (a)),
the generated data agree with the observed records in wider frequency range [0.1, 20] Hz.
This trend may be attributed to the influence of the generative processes of the SS-GMGM.
The constraint that the acceleration returns to zero under normal conditions is not included in the SS-GMGM. As a result, regions with small amplitudes periodically experience slight deviations from the zero line, which could result in low frequency noise.
As small magnitude earthquakes do not cause oscillations with low frequency components, the effect of low frequency noise is notably evident in the range of $5.5 \le M_W \le 5.7$ (Figure \ref{fig:fig_famp} (c)).
The frequency bands for which the distribution of the generated FAS agrees with the FAS of observed records are generally comparable to those examined in Florez et al. \cite{Florez2022} and Esfahani et al.\cite{esfahani2023tfcgan}.

\begin{figure}
  \centering
  \includegraphics[width=\columnwidth]{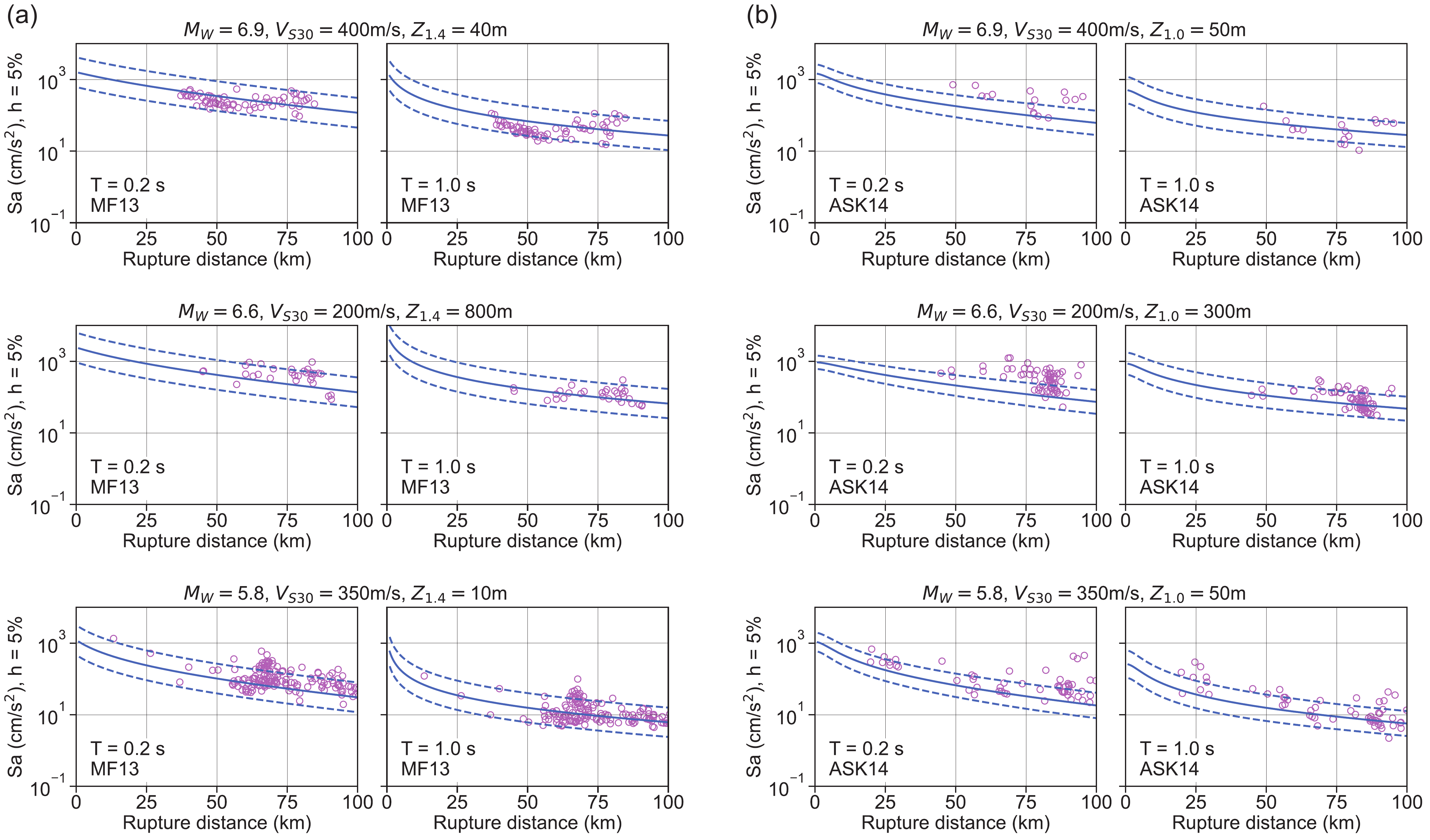}
  \caption{
    Comparison of the spectral acceleration values (5\% damping) at periods $T = 0.2$ and $1.0$ seconds between the generated ground motions and prediction results
    from the empirical GMMs. (a) represents the comparison results with the MF13 GMM across three different $M_W$, $R_{RUP}$, $V_{\mathrm{S}30}$, and $Z_{1.4}$ scenarios. (b) represents the comparison results with the ASK14 GMM across three different $M_W$, $R_{RUP}$, $V_{\mathrm{S}30}$, and $Z_{1.0}$ scenarios. 
    The purple circles represents the generated ground motions, the blue solid line represents the median of the GMMs prediction, and
    blue dashed line represents the $\pm 1$ standard deviation interval. The corresponding values of $M_W$ and soil conditions are shown at the top of each panel.
  }
  \label{fig:fig_sa_comp}
\end{figure}%

\subsection{Comparison with empirical GMMs}
The generated results of the SS-GMGM are compared with the prediction of the following two empirical GMMs:
\begin{enumerate}
  \item Morikawa and Fujiwara \cite{Morikawa2013} (MF13) GMM 
  \item Abrahamson et al. \cite{abrahamson2014summary} (ASK14) GMM 
\end{enumerate}

Morikawa and Fujiwara \cite{Morikawa2013} proposed two models which differ in amplitude scaling with reference to magnitudes.
We utilize a model with a quadratic magnitude term and perform predictions using the following formula:
\begin{align}
  \log_{10}S_a &= a\left(M_{W^\prime} - M_{W_1}\right)^2 + bR_{RUP} + c - \log_{10}\left(R_{RUP} + d\cdot 10^{eM_{W^\prime}}\right) + G_d + G_s \\
  M_{W^\prime} &= \min(M_{W}, M_{W_0}) \\
  G_z &= p_z\log_{10}\left(\frac{\max(Z_{\min}, Z_{1.4})}{Z_0}\right) \\
  G_s &= p_s\log_{10}\left(\frac{\min(V_{\mathrm{S}\max}, V_{\mathrm{S}30})}{V_0}\right)
\end{align}
where $S_a$ is the spectral acceleration value for specific period, $a$, $b$, $c$, $d$, $e$, $M_{W_0}$, $M_{W_1}$, $p_z$, $p_s$, $Z_{\min}$, and $V_{\mathrm{S}\max}$ are coefficients.
$G_z$ is the correction term for amplification by deep sedimentary layers, and $G_s$ is the correction term for amplification by shallow soft soils.

The ASK14 GMM contains some explanatory variables to specify the source conditions.
Since the SS-GMGM cannot take into account such detailed source conditions,
the following equation is used for the prediction, considering only terms related to site conditions and the region-specific term:
\begin{align}
  \log S_a &= f_1(M_W, R_{RUP}) + f_5(\widehat{Sa}_{1180}, V_{\mathrm{S}30}) + f_{10}(Z_{1.0}, V_{\mathrm{S}30}) + Regional(V_{\mathrm{S}30}, R_{RUP})
\end{align}
where $\widehat{Sa}_{1180}$ is the median spectral acceleration on hard rock.
For detailed formulation of $f_1$, $f_5$, $f_{10}$, and $Regional(\cdot)$, please refer to Abrahamson et al. \cite{abrahamson2014summary}.

\begin{figure}
  \centering
  \includegraphics[width=\columnwidth]{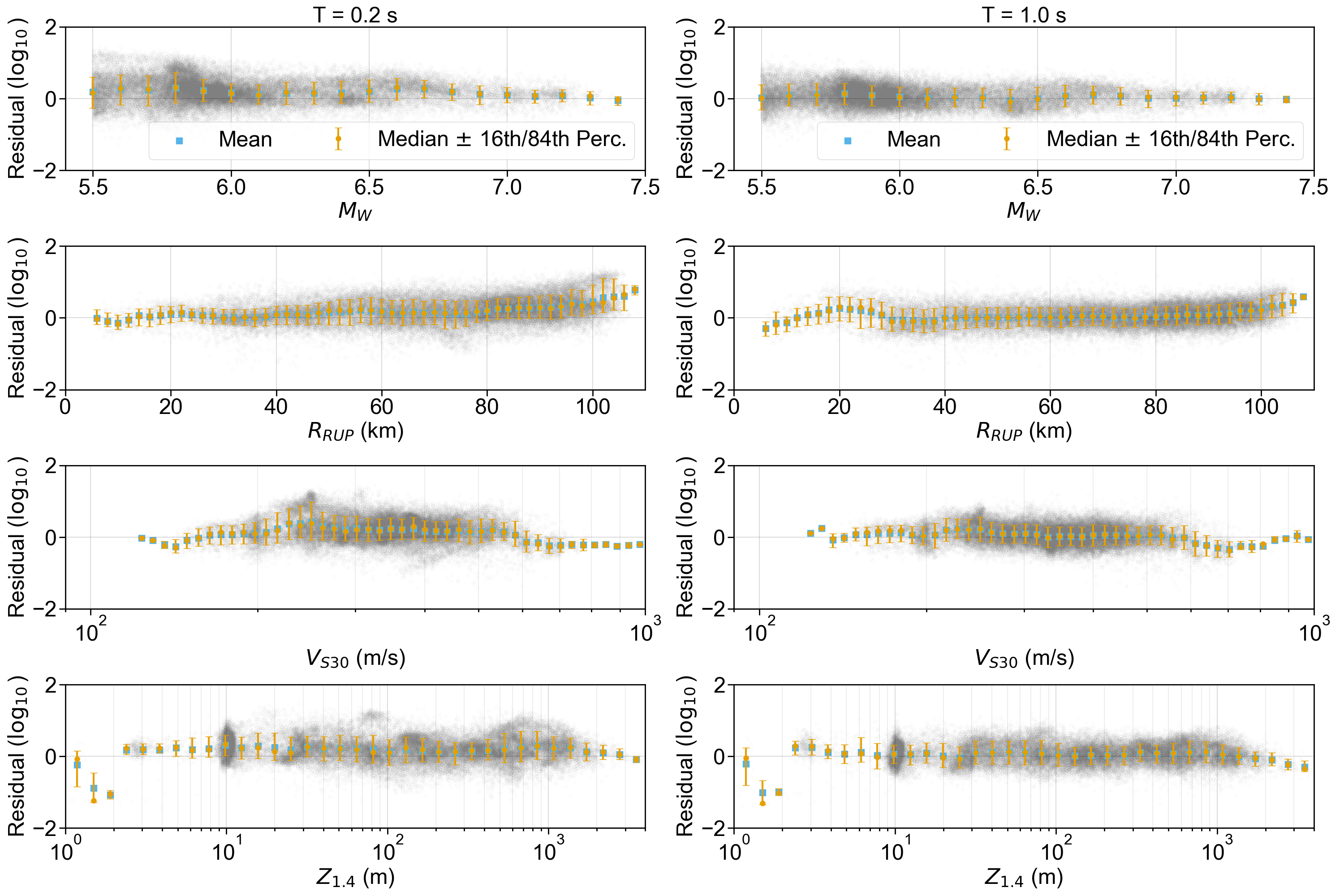}
  \caption{
    Residual plots between the spectral acceleration of ground motions generated by the SS-GMGM and the median predictions of the MF13 GMM.
    Grey circles represent the residuals for each generated data, blue squares indicate the mean, and orange bars show the median along with the 
    16th and 84th percentiles.
    Each panel represents the residuals for a period of $T = 0.2$ seconds on the left and $T = 1.0$ second on the right.
  }
  \label{fig:residual_sa_comp_mf14}
\end{figure}%

\subsubsection{Distribution}
Figure \ref{fig:fig_sa_comp} compares the spectral acceleration values (5\% damping) at periods $T = 0.2$ and $1.0$ seconds between the generated ground motions and prediction results from the MF13 GMM and ASK14 GMM. The distance scaling of the SS-GMGM is generally consistent with the MF13 GMM,
and the variability of generated data is also agree with the prediction of the MF13 GMM.
Compared to the median spectral acceleration values of the ASK14 GMM at $T = 0.2$ seconds, the values from the SS-GMGM tend to be slightly higher, especially under soft soil conditions. 
However, the distributions of the spectral acceleration values at $T = 1.0$ second are generally in agreement.

\begin{figure}
  \centering
  \includegraphics[width=\columnwidth]{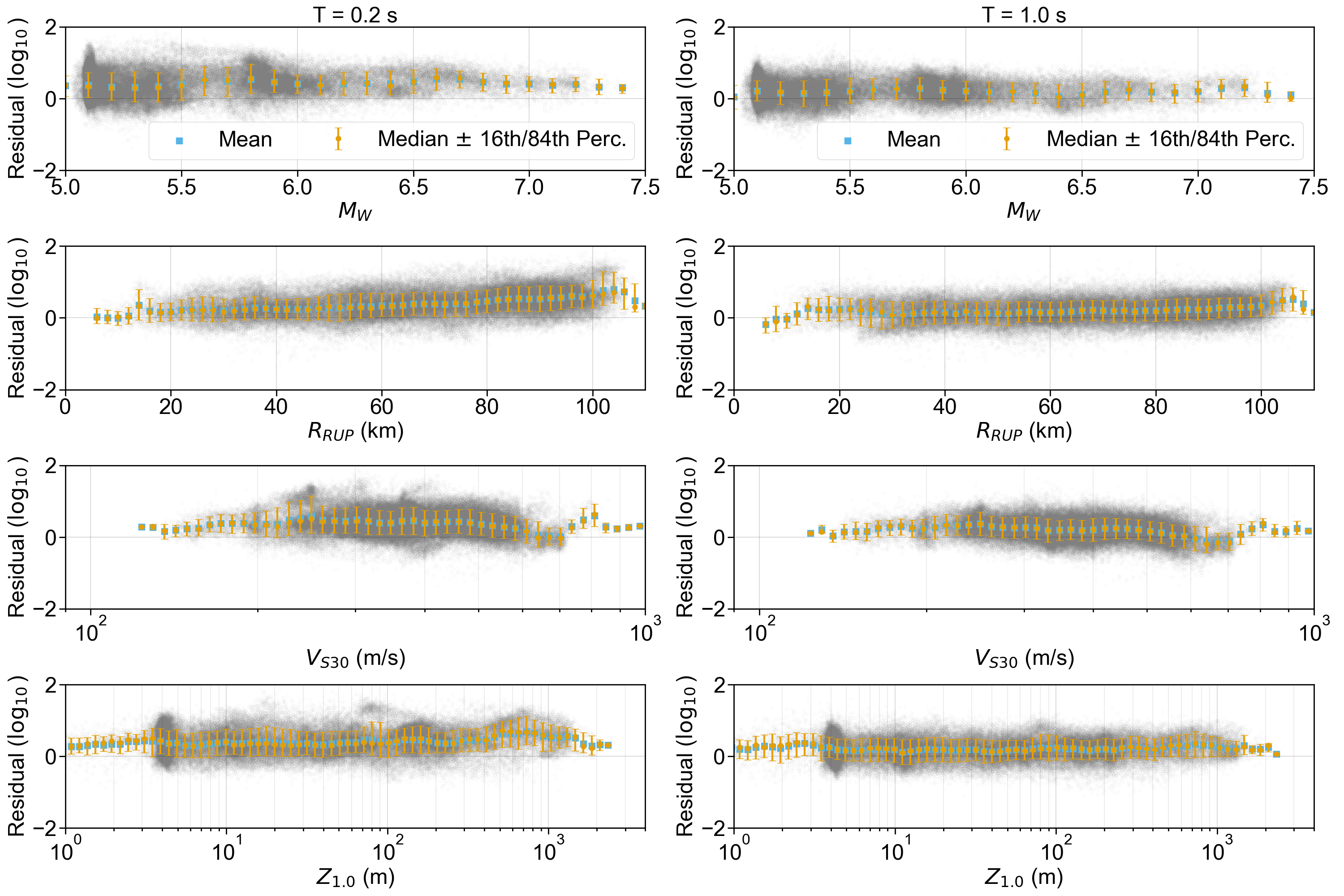}
  \caption{
    Residual plots between the spectral acceleration of ground motions generated by the SS-GMGM and the median predictions of the ASK14 GMM.
    Grey circles represent the residuals for each generated data, blue squares indicate the mean, and orange bars show the median along with the
    16th and 84th percentiles.
    Each panel represents the residuals for a period of $T = 0.2$ seconds on the left and $T = 1.0$ second on the right.
  }
  \label{fig:residual_sa_comp_ask14}
\end{figure}%

\subsubsection{Residual analysis}
The median values of the MF13 GMM and ASK14 GMM corresponding to the each generated condition label
are predicted, and the residual values ($\log_{10}[\text{gen}/\text{pre}]$, gen is the generated data and pre is the median of the GMMs) are calculated.
To align the condition of the dataset, only generate data with $M_W \ge 5.5$ is used for the MF13 GMM predictions.
Figure \ref{fig:residual_sa_comp_mf14} shows the residuals with reference to $M_W$, $R_{RUP}$, $V_{\mathrm{S}30}$, and $Z_{1.4}$ compared with the MF13 GMM.
The residuals are generally centered around zero regardless of $M_W$, $V_{\mathrm{S}30}$, and $Z_{1.4}$ values,
indicating that the SS-GMGM has appropriately learned the magnitude scaling as well as the amplification characteristics due to shallow soil and deep sedimentary layers.
In terms of distance scaling, there is a tendency to slightly overestimate in regions where $R_{RUP}$ is large,
and to slightly underestimate in regions with short distances at $T = 1.0$ seconds.
However, across a wide range, the residuals are generally centered around zero, showing that the distance scaling of the SS-GMGM is consistent with reasonable accuracy.
Figure \ref{fig:residual_sa_comp_ask14} shows the residual plots of the SS-GMGM against the ASK14 GMM prediction for each value of $M_W$, $R_{RUP}$, $V_{\mathrm{S}30}$, and $Z_{1.0}$.
The generated spectral accelerations tend to be slightly larger at $T = 0.2$ seconds, as also shown in Figure \ref{fig:fig_sa_comp}.
However, the mean values of the residuals remain relatively constant regardless of the values of $M_W$, $R_{RUP}$, $V_{\mathrm{S}30}$, and $Z_{1.0}$.
This confirms that the SS-GMGM has adequately learned the scaling of magnitude and distance, as well as of the amplifications by the shallow soil and deep sedimentary layers when compared with the ASK14 GMM.

\subsection{Evaluation as a site-specific model}\label{subsec:ss}
\begin{figure}
  \centering
  \includegraphics[width=\columnwidth]{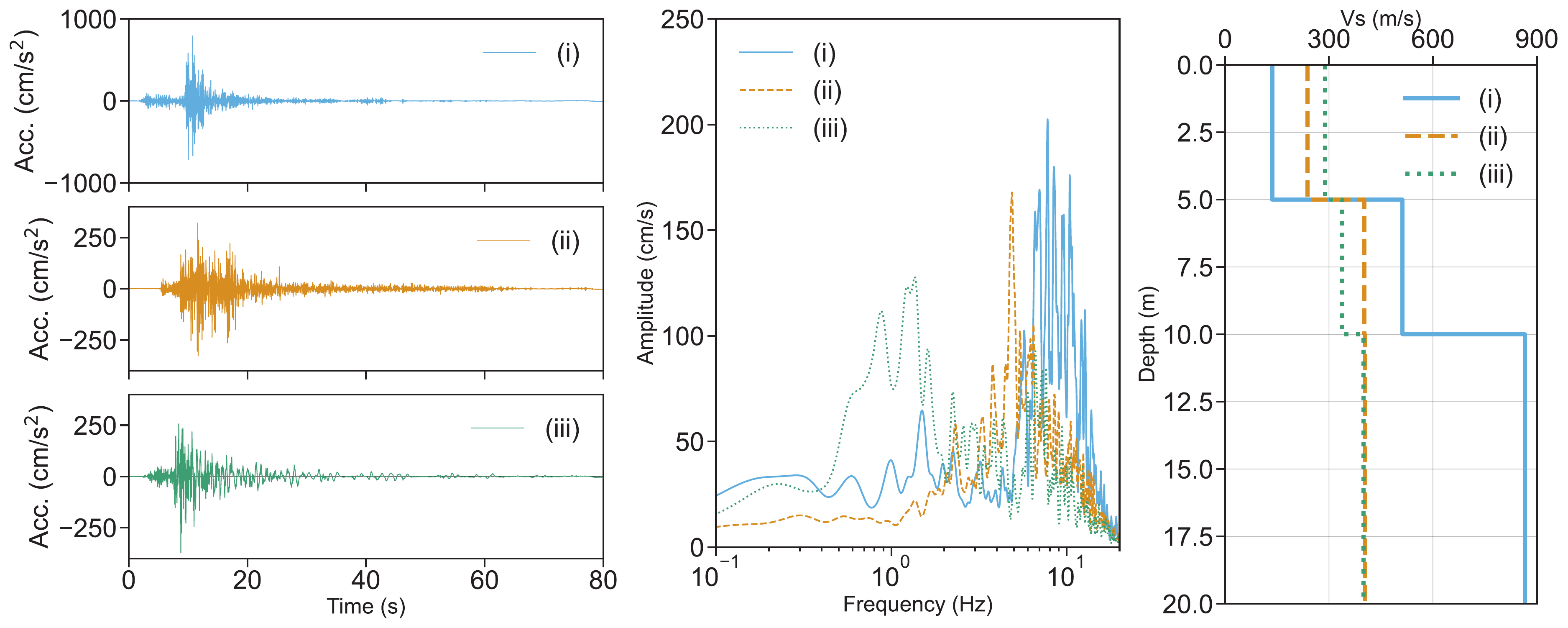}
  \caption{Examples of ground motion waveforms, FAS, shear-wave velocity profiles of generated data with almost same $M_W$, $R_{RUP}$, and $V_{\mathrm{S}30}$ values.
  }
  \label{fig:example_same_v30}
\end{figure}

The performance of the SS-GMGM is assessed in this subsection when site conditions are given in detail by looking at the relationship between generated ground motions and condition labels.
First, the shear-wave velocity profile modeled in three layers for the top 20 meters is back-calculated from the generated values of $V_{\mathrm{S}5}$, $V_{\mathrm{S}10}$, and $V_{\mathrm{S}20}$
as follows:
\begin{align}
  v_{0-5} &= V_{\mathrm{S}5} \notag \\
  v_{5-10} &= \frac{V_{\mathrm{S}5}V_{\mathrm{S}10}}{2V_{\mathrm{S}5} - V_{\mathrm{S}10}} \\
  v_{10-20} &= \frac{2v_{5-10}V_{\mathrm{S}5}V_{\mathrm{S}20}}{4v_{5-10}V_{\mathrm{S}5} - v_{5-10}V_{\mathrm{S}20} - V_{\mathrm{S}5}V_{\mathrm{S}20}} \notag
\end{align}
where $v_{0-5}$, $v_{5-10}$, and $v_{10-20}$ are the average shear-wave velocity of the first-layer (0-5 m), second-layer (5-10 m), and third-layer (10-20 m) of the surface soil, respectively.
Subsequently, generated data with nearly identical values of $M_W$, $R_{RUP}$, and $V_{\mathrm{S}30}$, which are used as condition labels of some GMGM (\cite{Florez2022}, \cite{esfahani2023tfcgan}), are sampled.
Figure \ref{fig:example_same_v30} presents the time-history waveforms, FAS, and corresponding velocity profiles for three generated samples,
and Table \ref{tab04} represents the corresponding condition label values.
Note that the FAS was smoothed using a Parzen window with a bandwidth of 0.2 Hz.
Even when the values of $M_W$, $R_{RUP}$, and $V_{\mathrm{S}30}$ are nearly the same, it is evident that the characteristics of the ground motions differ significantly.
In Figure \ref{fig:example_same_v30}, the generated ground motion in (i) has large amplitude, and has a peak near 8 Hz in FAS,
whereas the generated ground motion in (ii) has relatively smaller amplitude with a peak near 5 Hz in FAS.
Examining the corresponding shear-wave velocity profiles, the data for case (i) reveals a velocity profile where the first-layer has a shear-wave velocity of $v_{0-5} = 134$ m/s,
and the second-layer has a shear-wave velocity of $v_{0-5} = 512$ m/s.
This indicates a soft upper layer of 5 meters, underlain by a harder layer beyond 5 meters. Similarly, in case (ii),
the top 5 meters also consists of a soft layer, but $v_{0-5}$ is slightly larger than in (i), and the difference in shear-wave velocities between the first and second layer is smaller.
The high frequency vibrations are amplified due to multiple reflections at the soft surface soil,
suggesting that the generated data in (i) shows larger amplitude and higher predominant frequency compared to (ii) due to this amplification.

\begin{table}
  \caption{Values of the condition labels in Figure \ref{fig:example_same_v30}}
  \label{tab04}
  \begin{center}
    \begin{tabular}{cccc}
		\toprule
        \textbf{Condition label} & \textbf{(i)} & \textbf{(ii)} & \textbf{(iii)} \\
        \midrule
        {$M_W$} & 6.4 & 6.4 & 6.4 \\
        {$R_{RUP}$ (km)} & 34 & 32 & 34 \\
        {$V_{\mathrm{S}30}$ (m/s)} & 409 & 408 & 415 \\
        {$Z_{1.0}$ (m)} & 4.0 & 4.0 & 203 \\
        {$Z_{1.4}$ (m)} & 9.0 & 9.0 & 217 \\
        {$v_{0-5}$ (m/s)} & 134 & 238 & 289 \\
        {$v_{5-10}$ (m/s)} & 512 & 402 & 338 \\
        {$v_{10-20}$ (m/s)} & 866 & 404 & 400 \\
        \bottomrule
    \end{tabular}
  \end{center}
\end{table}

Furthermore, the data in (iii), while having similar amplitude and shear-wave velocity profile compared to (ii), contain many low frequency components.
The $Z_{1.0}$ and $Z_{1.4}$ values for (iii) are relatively high,
indicating that the generated ground motions are also consistent with the effect of deep sedimentary layers.
The SS-GMGM is capable of representing characteristic of ground motions that cannot be explained solely by $V_{\mathrm{S}30}$ value.

\begin{figure}
  \centering
  \includegraphics[width=0.7\columnwidth]{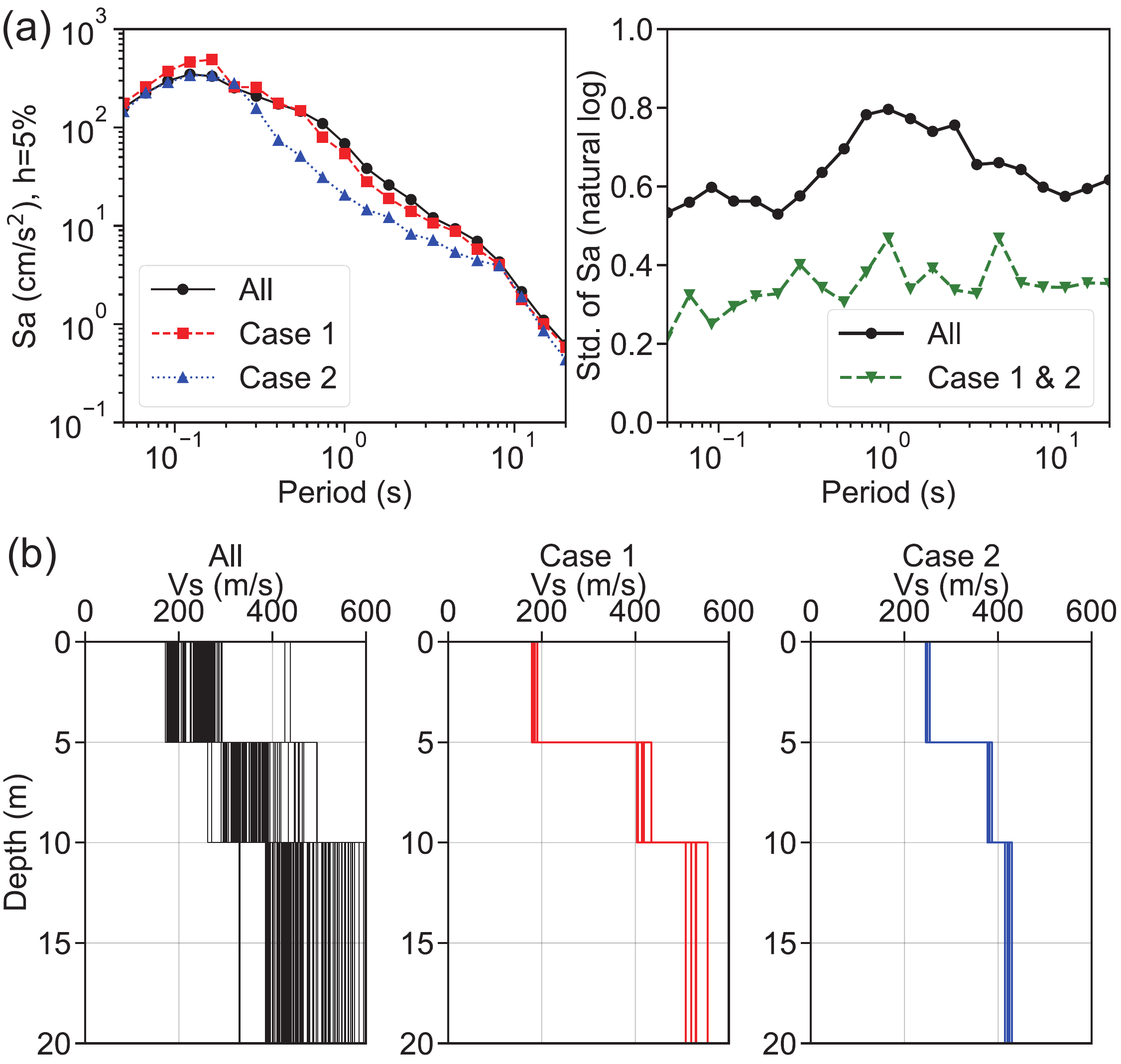}
  \caption{
    The distribution of the generated ground motions with some magnitude, distance, and site conditions scenarios.
    Panel (a) represents the means and standard deviations (in natural log units) of 5\% damped acceleration spectra for three scenarios.
    The black solid line with circles represent the distribution of generated acceleration spectra with specified values of $M_W$, $R_{RUP}$, and $V_{\mathrm{S}30}$.
    The red dashed line with squares (Case 1) and blue dotted line with triangles (Case 2) indicate the mean acceleration spectra of generated ground motions,
    which include additional specifications of the shear-wave velocity profiles along with the aforementioned $M_W$, $R_{RUP}$, and $V_{\mathrm{S}30}$. 
    The standard deviations for Case 1 and Case 2 are calculated together to ensure a sufficient sample size and are represented by a green dashed line with inverted triangles.
    Panel (b) represents the corresponding shear-wave velocity profiles for each generated data of the three scenarios.
  }
  \label{fig:main_sa_dist_comp}
\end{figure}%

\begin{table}
  \caption{Condition label values of the six generated ground motions of Case 2 in Figure \ref{fig:main_sa_dist_comp} and site condition of YMG007 station.}
  \label{tab05}
  \begin{center}
    \begin{tabular}{cccccccc}
		\toprule
        \textbf{Condition labels} & \textbf{(1)} & \textbf{(2)} & \textbf{(3)} & \textbf{(4)} & \textbf{(5)} & \textbf{(6)} & \textbf{YMG007}\\
        \midrule
  {$M_W$} & 6.0 & 6.0 & 6.0 & 6.0 & 6.0 & 6.0 & - \\
  {$R_{RUP}$ (km)} & 29 & 33 & 31 & 32 & 33 & 35 & - \\
  {$V_{\mathrm{S}30}$ (m/s)} & 416 & 411 & 418 & 417 & 420 & 416 & 418 \\
  {$v_{0-5}$ (m/s)} & 249 & 246 & 254 & 248 & 249 & 247 & 260 \\
  {$v_{5-10}$ (m/s)} & 377 & 378 & 380 & 387 & 382 & 378 & 400 \\
  {$v_{10-20}$ (m/s)} & 422 & 414 & 420 & 425 & 430 & 424 & 400 \\
  {$Z_{1.0}$ (m)} & 3.9 & 4.2 & 4.0 & 4.0 & 3.9 & 3.9 & 4.0 \\
  {$Z_{1.4}$ (m)} & 10 & 11 & 10 & 10 & 10 & 10 & 10 \\
        \bottomrule
    \end{tabular}
  \end{center}
\end{table}

\begin{figure}[t]
  \centering
  \includegraphics[width=0.7\columnwidth]{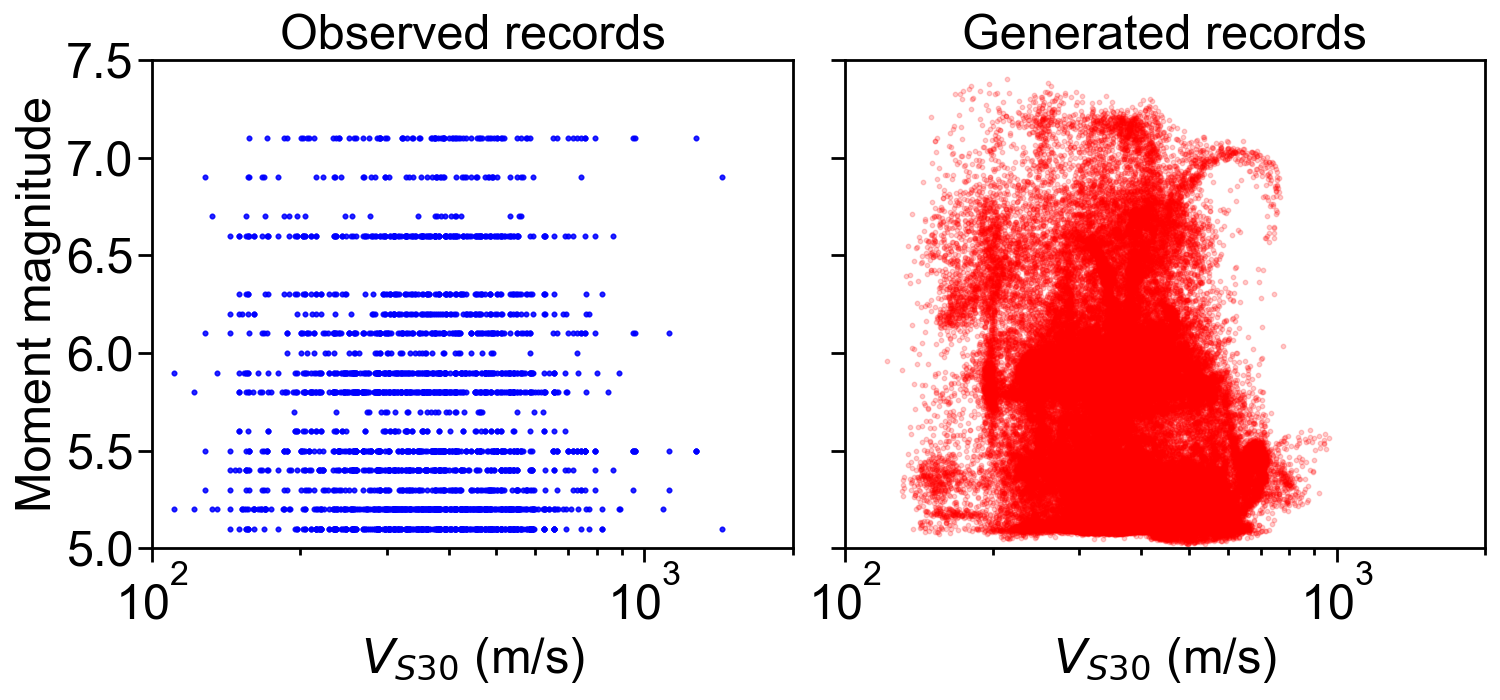}
  \caption{Comparison of the magnitude-$V_{\mathrm{S}30}$ distributions of generated data with observed records database.}
  \label{fig:comp_vs30_mw_dist}
\end{figure}%

\subsection{Distribution of generated ground motions with specified magnitude, distance, and detailed site conditions}
We examine the distribution of generated ground motions when magnitude, distance, and detailed site conditions are specified.
Initially, generated data fitting the following criteria were sampled, and 131 data are obtained:
\begin{itemize}
  \item $5.9 \le M_W \le 6.1$
  \item $25\:\text{km}\le R_{RUP} \le 35\:\text{km}$
  \item $404\:\text{m/s} \le V_{\mathrm{S}30} \le 426\:\text{m/s}$
\end{itemize}
The black dashed line with circles in Figure \ref{fig:main_sa_dist_comp} (a) shows the mean and standard deviation of the acceleration spectra for all 131 sampled generated ground motions.
Among the 131 samples, we further extracted two clusters that share similar shear-wave velocity profiles (Cases 1 and 2).
The red dashed line with squares (Case 1) and blue dotted line with triangles (Case 2) in Figure \ref{fig:main_sa_dist_comp} (a)
shows the mean acceleration spectra of the extracted clusters, and their corresponding shear-wave velocity profiles of each sampled data
are shown in Figure \ref{fig:main_sa_dist_comp} (b).
Note that the standard deviation of Case 1 and Case 2 are calculated together to ensure a sufficient sample size as follows:
\begin{equation}
  \sigma = \sqrt{\frac{(n_1 - 1)\hat{\sigma}_1^2 + (n_2 - 1)\hat{\sigma}_2^2}{n_1 + n_2 - 2}}
\end{equation}
where $n_1$ and $n_2$ are the number of data in Case 1 and 2, respectively, and $\hat{\sigma}_1^2$ and $\hat{\sigma}_2^2$ are the unbiased variance of data in Case 1 and 2, respectively.
The mean acceleration spectra varies depending on the soil profile.
For example, Case 1 has relatively soft soil condition of top 5 meters, and the spectral acceleration of Case 1 at short period range had larger values.
The standard deviations at each period have decreased by specifying the shear-wave velocity profile,
and the averaged standard deviation for each period of Case 1 and Case 2 results in about 0.35.
According to the study by Morikawa et al. \cite{morikawa2008strong}, the standard deviation (total of the within-event and between-events) of the 5\% damped spectral acceleration at specific station
generally ranges between 0.35 and 0.45.
Hikita and Tomozawa \cite{Hikita2023} studied the variability of spectral acceleration of single-path ground motions,
and the standard deviations approximately ranged between 0.3 and 0.5 in their examination.
Although a strict comparison is challenging due to different settings, the standard deviation obtained in this study is considered to be generally reasonable.

\subsection{Data interpolation by SS-GMGM}
GANs are known to generate new data samples by interpolating the distribution of a learned dataset \cite{luzi2021double}.
In this subsection, we evaluate the data interpolation of the SS-GMGM for generated ground motions and condition labels.

First, Figure \ref{fig:comp_vs30_mw_dist} compares the distribution of $M_W$ and $V_{\mathrm{S}30}$ between the observed records dataset and generated data.
While no generated data correspond to the slightly included area of $V_{\mathrm{S}30} > 1000$ m/s in the training dataset,
new data have been generated within the region where observed records are distributed.
For instance, there are no observed records corresponding to $M_W = 6.5$ in the dataset, 
yet the SS-GMGM also generates data for such regions.
Looking at the residual plots in Figure \ref{fig:residual_sa_comp_mf14}, the generated results for $M_W = 6.5$ are consistent with the predictions of the MF13 GMM, indicating that the generated ground motions are reasonable.
The SS-GMGM can generate data for combinations of condition labels that are not included in the observed record database.

Next, we examine the case specifying $M_W$, $R_{RUP}$, and the shear-wave velocity profile.
Here, we use the generated results of Case 2 shown in Figure \ref{fig:main_sa_dist_comp} as an example.
Table \ref{tab05} lists the values of the condition labels of six generated ground motions in Case 2.
For these generated data, the values for $Z_{1.0}$ and $Z_{1.4}$, which were not specified during data sampling,
are almost identical.
Similar trends were occasionally observed when different combinations of the $M_W$, $R_{RUP}$, and shear-wave velocity profile were specified.
Table \ref{tab05} also shows the site conditions of YMG007 station, a K-NET station included in our training dataset.
It is important to note that since only shear-wave velocity profile up to 10 meters has been obtained at YMG007 station in K-NET database,
it is assumed that the shear-wave velocity from 10 meters to 20 meters is identical to the upper layer.
The shear-wave velocity profile and values of $Z_{1.0}$ and $Z_{1.4}$ at YMG007 station are very similar to those of the generated data in Case 2.
This suggests that while the SS-GMGM can interpolate data when only a few condition labels are considered, as shown in Figure \ref{fig:comp_vs30_mw_dist},
such interpolation becomes more challenging as the number of considered condition labels increases.
This will be discussed further in section \ref{sec:conc}.

\begin{figure}[t]
  \centering
  \includegraphics[width=0.6\columnwidth]{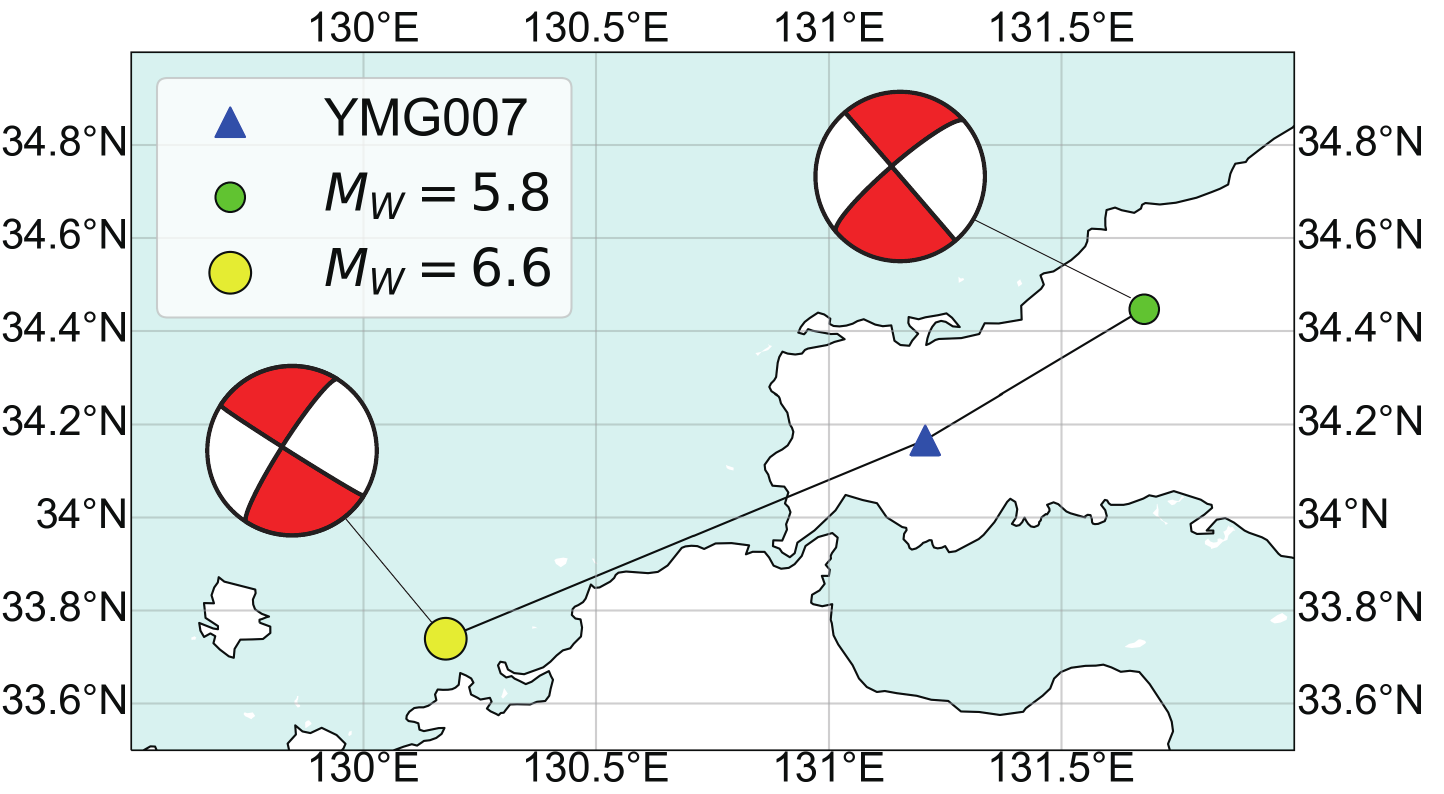}
  \caption{The locations of the YMG007 K-NET station (triangle) and earthquake epicenters (circles) whose ground motions were observed at YMG007 station and contained in our training dataset.
  The focal mechanisms were obtained from F-net database.}
  \label{fig:station_map_ymg007}
\end{figure}%

\begin{figure}[t]
  \centering
  \includegraphics[width=0.5\columnwidth]{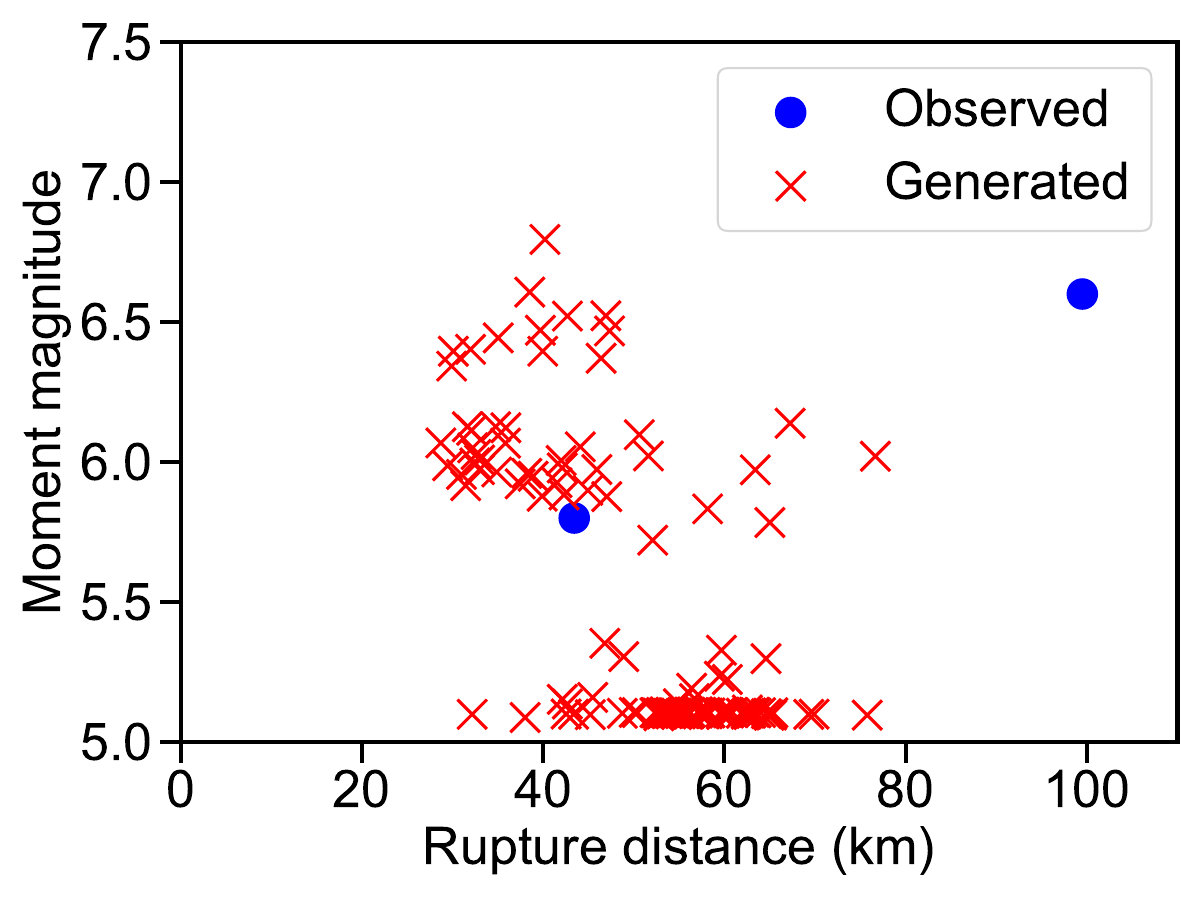}
  \caption{Magnitude-distance distribution of the observed records at YMG007 K-NET station (blue circle) and generated ground motions (red cross) with
  site conditions similar to YMG007 station.}
  \label{fig:scat_ymg007}
\end{figure}%

\begin{figure}[t]
  \centering
  \includegraphics[width=0.51\columnwidth]{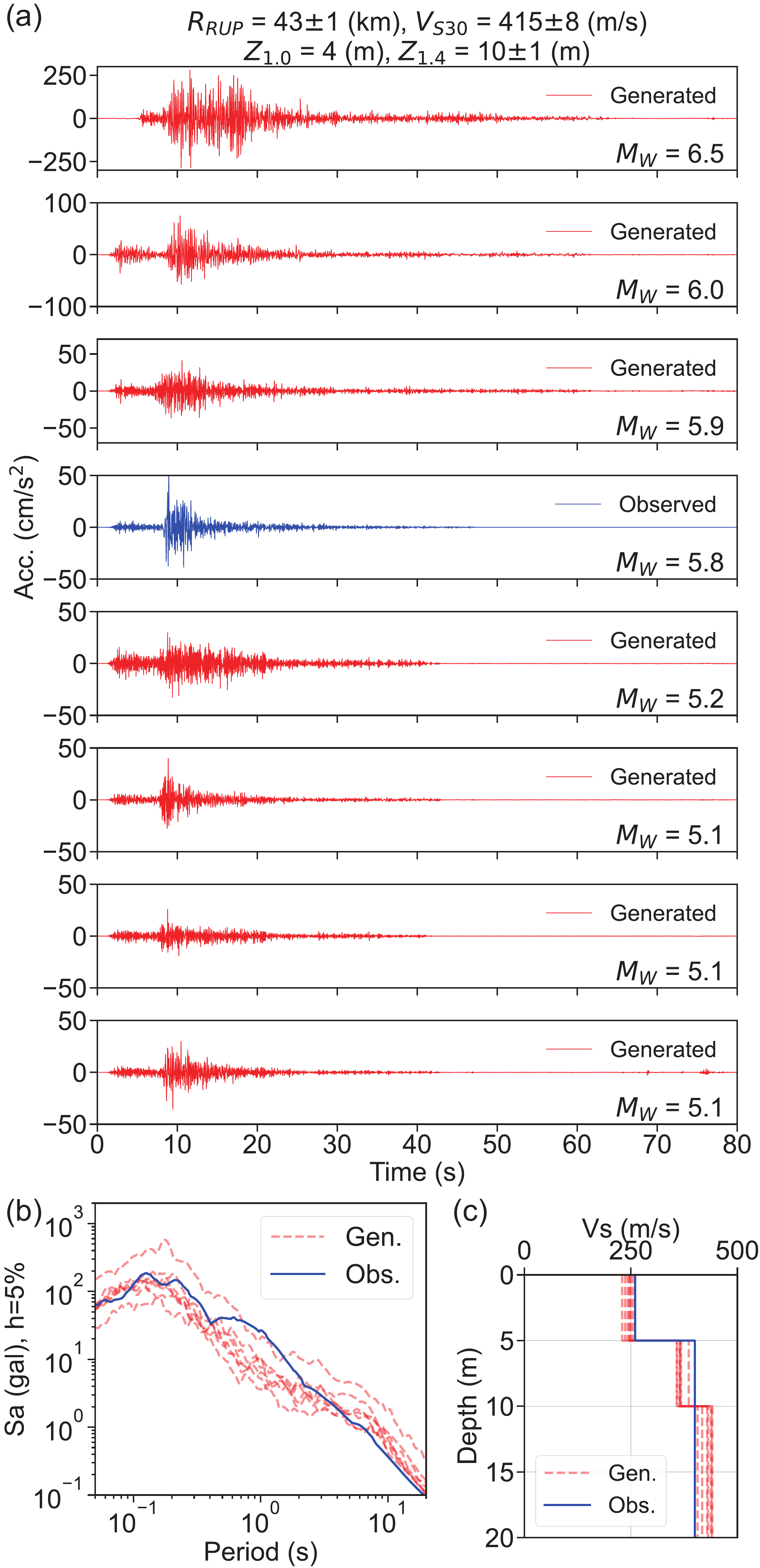}
  \caption{Ground motion waveforms (a) and their corresponding acceleration spectra (b) and shear-wave velocity profiles (c) that satisfy the condition $42\:\text{km} \le R_{RUP} \le 44$ km in Figure \ref{fig:scat_ymg007}.
  The blue solid line represents the observed record, and red dashed line represent the generated data.
  All condition label values for the data shown in this figure are distributed within the range indicated at the top of the figure.
  }
  \label{fig:comp_ymg_wave_437}
\end{figure}

Finally, we examine the case where only site conditions were specified.
At YMG007 station, two observed records from different earthquakes are obtained in our training dataset.
The relationships between these earthquakes and the YMG007 station are shown in
Figure \ref{fig:station_map_ymg007}.
Then, generated data having the same site conditions as YMG007 station are sampled, corresponding to the following criteria:
\begin{itemize}
\item $225\:\text{m/s} \le v_{0-5} \le 275\:\text{m/s}$
\item $355\:\text{m/s} \le v_{5-10} \le 405\:\text{m/s}$
\item $395\:\text{m/s} \le v_{10-20} \le 445\:\text{m/s}$
\item $3\:\text{m} \le Z_{1.0} \le 5\:\text{m}$
\item $9\:\text{m} \le Z_{1.4} \le 11\:\text{m}$
\end{itemize}
Figure \ref{fig:scat_ymg007} shows the magnitude-distance distribution of the two observed records at YMG007 station and the sampled generated data.
Although no data have been generated for region where $R_{RUP}$ is approximately 100 km,
it is evident that data corresponding to combinations of magnitude and distance not included in the observed records have been generated even when specifying detailed site conditions.
Therefore, even when the number of considered condition labels is increased, the SS-GMGM is capable of generating data not included in the observed records dataset, provided that the number of condition labels used for interpolation remains small.
For the data for $42\:\text{km} \le R_{RUP} \le 44$ km in Figure \ref{fig:scat_ymg007},
time history waveforms and their corresponding acceleration spectra and shear-wave velocity profiles are shown in Figure \ref{fig:comp_ymg_wave_437}.
The frequency characteristics of generated ground motions are generally similar to the observed record,
and the magnitude scaling of the amplitude is generally appropriate.

\section{Conclusion and Discussion}\label{sec:conc}
In this study, we developed a GMM (SS-GMGM) for crustal earthquakes in Japan that can directly generate ground motion time histories with specifying detailed site conditions.
The proposed SS-GMGM were developed based on a type of deep generative model, called styleGAN2, and a novel neural network architecture that could serve as
a generative model with detailed condition labels were also proposed.
The neural network architecture of the SS-GMGM is capable of account for site conditions with five values, $[V_{\mathrm{S}5}, V_{\mathrm{S}10}, V_{\mathrm{S}20}, Z_{1.0}, Z_{1.4}]$, in addition to magnitude and distance information.
We demonstrated that the characteristics of the generated ground motions are consistent with these condition labels,
and ground motions with engineering-significant characteristic can be generated.
Furthermore, the amplifications by shallow soil and deep sedimentary layer were shown to be accurately represented, and the SSGMGM's magnitude and distance scaling were shown to match those of empirical GMMs.
Additionally, by modeling the surface soil in three layers, the SS-GMGM could express differences in ground motion characteristics that could not be explained solely by $V_{\mathrm{S}30}$,
and generated new data samples with different magnitude-distance settings even when detailed site conditions were specified.

\begin{figure}[t]
  \centering
  \includegraphics[width=0.5\columnwidth]{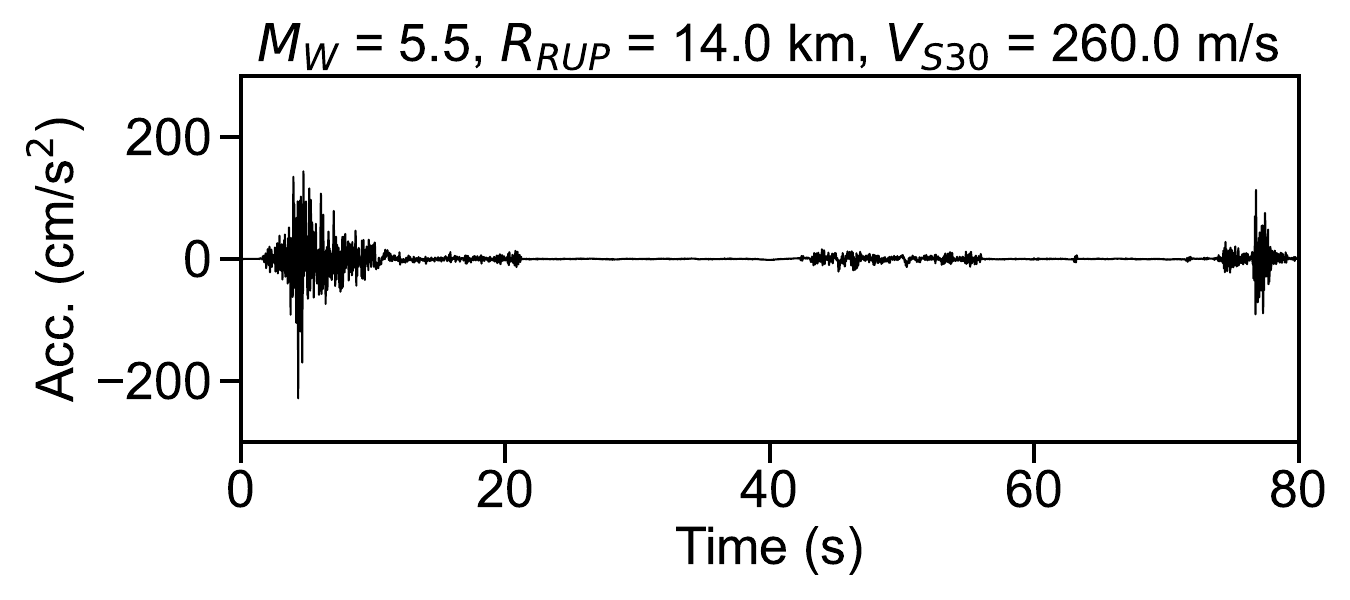}
  \caption{An example of generated ground motion waveform that subsequent oscillations showed large amplitude noise.
  The values at the top of this figure represent the corresponding condition label values.}
  \label{fig:gen_wave_ff}
\end{figure}%

The SS-GMGM constructed in this study does not always produce high-quality ground motions, in some cases, significant noises occur at the end of the data.
Figure \ref{fig:gen_wave_ff} shows an example of such generated result.
This type of data tends to be generated, although infrequently, under combinations of condition labels where there are fewer data points in training dataset.
The cause of this tendency might be attributed to the influence of the vibrations included in the observed records.
Subsequent vibrations believed to be caused by aftershocks have been eliminated, as stated in the data preprocessing subsection. On the other hand,
aftershocks left some vibrations that were difficult to pinpoint.
Therefore, the observed records in our training dataset might include vibrations like those in the time interval from 40 to 60 seconds in Figure \ref{fig:gen_wave_ff},
and it is conceivable that the SS-GMGM has learned such tendencies.
The vibrations seen after about 75 seconds in Figure \ref{fig:gen_wave_ff} are not mere noise but resemble a ground motion.
This could be due to the characteristics of convolutional neural networks (CNNs).
The parts of neural networks of the SS-GMGM have constructed by CNNs,
which are known to have a property called equivariance to translation due to parameter sharing \cite{Goodfellow-et-al-2016}.
It is important to improve the performance of the SS-GMGM by carefully examining the data preprocessing procedures and the
configuration of the neural networks suitable for generating ground motion data.

GANs generate new data samples by interpolating the distribution of learned dataset.
Although interpolation could be performed when a few condition labels are specified,
the site conditions specified by the five-dimensional vector $[V_{\mathrm{S}5}, V_{\mathrm{S}10}, V_{\mathrm{S}20}, Z_{1.0}, Z_{1.4}]$
were not well-interpolated.
This issue might stem from the {\it curse of dimensionality}.
In the compiled dataset, there are 728 observation stations.
The distribution of soil conditions defined by five-dimensional vectors is relatively high-dimensional considering this number of stations.
Data points may not be sufficiently dense for effective interpolation when many conditions are specified.
This challenge is one of the difficulties in data-driven approaches.
As more condition labels are considered in detail for the source, path, and site conditions,
the lack of data becomes a more significant issue.
Potential solutions to this problem about site conditions include increasing the number of observation stations used in compiled dataset.
The performance of the SS-GMGM may be improved by including observed records at KiK-net stations, as they were not used in this investigation. Using simulation-based techniques to estimate ground motion amplifications and predict ground motions at the bedrock using GMGMs could be another way to solve this problem.
This approach would require detailed site condition information, but would allow the avoidance of issues mentioned above in constructing the SS-GMGM.

Previous studies applying GANs to GMM (\cite{Florez2022}, \cite{esfahani2023tfcgan}, \cite{shi2024broadband}) have constructed models based on cGAN,
specifying condition labels for generating ground motions.
In contrast, this study generates ground motions and condition labels from normal random inputs without specifying such labels.
The advantages and disadvantages of our method compared to cGAN based method are as follows. 
An advantage is its suitability for generating ground motions when detailed condition labels are specified.
As mentioned earlier, GANs approximate the distribution of the learned dataset, making them unsuitable for generating data that would be a complete extrapolation from the dataset range.
For example, in the dataset used in this study, the moment magnitudes of the records range from 5.0 to 7.1, making it difficult for the SS-GMGM to generate ground motions for magnitudes like 4 or 8.
The distribution of condition labels within the seven-dimensional space is sparse, considering the number of available observed records.
Since cGAN requires the specification of condition labels to generate data, such sparsity could make it challenging to appropriately set these labels without exceeding the applicable range.
Consequently, models that specify more detailed condition labels may generate ground motions that ideally should not be generated.
Such problems can be avoided by generating condition labels with ground motions at the same time.
On the other hand, a disadvantage of our method is the time required to generate ground motions that matches specific condition labels.
The frequency of generated data depends on its occurrence in the training dataset,
which means data for less common conditions is less likely to be generated.
To utilize the SS-GMGM in the filed of earthquake engineering, it is necessary to construct a framework that incorporates rare-event simulation techniques \cite{bucklew2004introduction}, which are studied in fields such as structural reliability \cite{AU2001263}.

Additionally, while this study constructed GMGMs for the horizontal one component of ground motions occurred in crustal earthquakes,
it is also important from the viewpoint of earthquake engineering to predict ground motions for subduction zone earthquakes and
to simultaneously predict three components.
Expanding the application range of the SS-GMGM is essential and is a future task.

\section{Data and Resources}
  The strong motion records and the shear-wave velocity values of the surface soil
  used in this study can be downloaded through the website of the National Research Institute for Earth Science and Disaster Resilience
  (NIED; \url{https://www.kyoshin.bosai.go.jp/kyoshin/}). We only used the records of K-NET.
  The moment magnitude was obtained from the NIED F-net database \url{https://www.fnet.bosai.go.jp/},
  and the values of $Z_{1.0}$ and $Z_{1.4}$ were obtained from the NIED J-SHIS website \url{https://www.j-shis.bosai.go.jp/}.
  The program code used in deep learning is available in the GitHub repository \url{https://github.com/Mat-main-00/ss_gmgm}.

\section{Declaration of Competing Interests}
The authors acknowledge that there are no conflicts of interest recorded.

\section{Acknowledgments}
This study was supported by JSPS KAKENHI Grant Numbers JP23H00219 and JP22J23006.

\bibliographystyle{unsrt}

\end{document}